\documentclass[aps,prl,longbibliography,showpacs,twocolumn,superscriptaddress]{revtex4-2}
\usepackage{amsmath,amssymb,amsfonts,bm}
\usepackage{booktabs}
\usepackage{braket}
\usepackage{graphicx}
\usepackage{epstopdf}
\usepackage{dcolumn}
\usepackage{mathrsfs}
\usepackage{bbold}
\usepackage{dsfont}
\usepackage{float}
\usepackage{soul}
\usepackage[colorlinks=true,linkcolor=blue,citecolor=blue, urlcolor=blue,bookmarks=false]{hyperref}
\usepackage{amsbsy}
\usepackage{stmaryrd}
\usepackage{babel}
\usepackage{tgtermes}
\usepackage{multirow}
\usepackage[title,titletoc]{appendix}

\begin{document}

\title{Light-induced Odd-parity Magnetism in Conventional  Antiferromagnetism}

\author{Shengpu Huang}
\affiliation{Institute for Structure and Function $\&$ Department of Physics $\&$ Chongqing Key Laboratory for Strongly Coupled Physics, Chongqing University, Chongqing 400044, People's Republic of China}
\affiliation{Center of Quantum materials and devices, Chongqing University, Chongqing 400044, People's Republic of China}

\author{Zheng Qin}
\affiliation{Institute for Structure and Function $\&$ Department of Physics $\&$ Chongqing Key Laboratory for Strongly Coupled Physics, Chongqing University, Chongqing 400044, People's Republic of China}
\affiliation{Center of Quantum materials and devices, Chongqing University, Chongqing 400044, People's Republic of China}

\author{Fangyang Zhan}
\affiliation{Institute for Structure and Function $\&$ Department of Physics $\&$ Chongqing Key Laboratory for Strongly Coupled Physics, Chongqing University, Chongqing 400044, People's Republic of China}
\affiliation{Center of Quantum materials and devices, Chongqing University, Chongqing 400044, People's Republic of China}

\author{Dong-Hui Xu}
\email{donghuixu@cqu.edu.cn}
\affiliation{Institute for Structure and Function $\&$ Department of Physics $\&$ Chongqing Key Laboratory for Strongly Coupled Physics, Chongqing University, Chongqing 400044, People's Republic of China}
\affiliation{Center of Quantum materials and devices, Chongqing University, Chongqing 400044, People's Republic of China}

\author{Da-Shuai Ma}
\email{mads@cqu.edu.cn}
\affiliation{Institute for Structure and Function $\&$ Department of Physics $\&$ Chongqing Key Laboratory for Strongly Coupled Physics, Chongqing University, Chongqing 400044, People's Republic of China}
\affiliation{Center of Quantum materials and devices, Chongqing University, Chongqing 400044, People's Republic of China}

\author{Rui Wang}
\email[]{rcwang@cqu.edu.cn}
\affiliation{Institute for Structure and Function $\&$ Department of Physics $\&$ Chongqing Key Laboratory for Strongly Coupled Physics, Chongqing University, Chongqing 400044, People's Republic of China}
\affiliation{Center of Quantum materials and devices, Chongqing University, Chongqing 400044, People's Republic of China}

\begin{abstract}
Recent studies have drawn growing attention on non-relativistic odd-parity magnetism in the wake of altermagnets. Nevertheless, odd-parity spin splitting is often believed to appear in non-collinear magnetic configurations. Here, using symmetry arguments and effective model analysis, we show that Floquet engineering offers a universal strategy for achieving odd-parity magnetism in two-dimensional (2D) collinear antiferromagnetism under irradiation of periodic driving light fields such as circularly polarized light, elliptically polarized light, and bicircular light.  The symmetry requirements and three distinct lattice models of potential candidates are established. 
Strikingly, the light-induced odd-parity spin splitting can be flexibly controlled by adjusting the crystalline symmetry or the polarization state of incident light, enabling the reversal or conversion of spin-splitting. By combining first-principles calculations and Floquet theorem, we present illustrative examples of 2D collinear antiferromagnetic (AFM) materials to verify the light-induced odd-parity magnetism. Our work not only offers a powerful approach for uniquely achieving odd-parity spin-splitting with high tunability, but also expands the potential of Floquet engineering in designing unconventional compensated magnetism. 
\end{abstract}
\maketitle

\textit{\textcolor{blue}{Introduction.}}---
Recently, a novel category of collinear antiferromagnetic (AFM) phase dubbed altermagnetism, characterized by fully symmetry-compensated magnetic moments in real space and non-relativistic spin splitting in reciprocal space, has become the subject of intense recent studies~\cite{PhysRevB.102.144441,RuO22022PhysRevLett.128.197202,PhysRevX.12.040002,WCJ2007PhysRevB.75.115103,PhysRevLett.130.036702,hayami2019momentum,PhysRevLett.132.176702,MnTe2024krempasky2024altermagnetic,yao2024bai2024altermagnetism,song2025altermagnets,cheong2025altermagnetism,zhang2025crystal,PhysRevX.15.021083}. The symmetry of altermagnets is dictated by spin space groups, which establish a new paradigm for understanding spin-dependent phenomena in crystalline materials~\cite{brinkman1966theory,spingroup12022PhysRevX.12.021016,LB2022PhysRevX.12.040501,spingroup22024PhysRevX.14.031037,PhysRevX.14.031039}. 
Combining the advantages of ferromagnetism and antiferromagnetism in
terms of spin splitting, altermagnets are predicted to exhibit a variety of fascinating properties, such as crystal Hall effects~\cite{vsmejkal2020crystal}, spin-dependent conductivity~\cite{LB2022PhysRevX.12.031042,LB2022PhysRevX.12.040501}, large magnetoresistance effects~\cite{MR2022PhysRevX.12.011028,shao2021spin}, as well as beyond~\cite{OP2021PhysRevB.104.024401,PM2021ma2021multifunctional,PM2024PhysRevMaterials.8.L041402,reichlova2024observation,PhysRevLett.130.046401,PhysRevLett.133.166701,PhysRevLett.133.206702,PhysRevB.110.174410,PhysRevLett.134.166701,PhysRevB.108.075425,PhysRevLett.131.076003,PhysRevB.108.184505,PhysRevLett.134.106802,zm5y-vy41,PhysRevLett.134.106801}. 
To date, altermagnetism has been theoretically predicted and partially confirmed in a wide range of materials, stimulating broad interest in both fundamental studies and potential device applications~\cite{song2025altermagnets,sodequist2024two,10.1093/nsr/nwaf066,PhysRevMaterials.9.064403}.

It is worth noting that altermagnetism, in its initial definition, features even-parity spin splitting with symmetries such as $d$-wave, $g$-wave, and $i$-wave~\cite{LB2022PhysRevX.12.040501,LB2022PhysRevX.12.031042}.
Following in the footsteps of counterpart altermagnets, non-relativistic odd-parity magnets have emerged as a prominent and rapidly evolving research frontier~\cite{hellenes2024pwavemagnets,oddNatruesong2025electrical,yu2025oddparitymagnetismdrivenantiferromagnetic,PhysRevLett.133.236703,PhysRevB.111.125420,lin2025odd,yamada2025metallic}. The odd-parity magnetism exhibits spin-splitting that is odd under a sign change of the momentum, analogous to the well-established Rashba and Dresselhaus spin-orbital coupling, suggesting its promising potential for spintronic devices~\cite{RevModPhys.76.323,zhang2014hidden,manchon2015new,koo2020rashba}. Among various odd-parity magnets, $p$-wave magnetism is of particular interest as it corresponds to the long-sought unconventional $p$-wave superfluidity~\cite{RevModPhys.47.331,RevModPhys.72.969,RevModPhys.75.657}, and significant theoretical and experimental advances have been made in this field~\cite{hellenes2024pwavemagnets,oddNatruesong2025electrical,PhysRevLett.133.236703}.
In particular, most recently, experimental evidence of $p$-wave spin splitting and its electrical control has been observed in NiI$_2$~\cite{oddNatruesong2025electrical}.
Despite recent encouraging advancements, odd-parity spin splitting reported to date has been limited to non-collinear magnetic configurations~\cite{hellenes2024pwavemagnets,oddNatruesong2025electrical,yu2025oddparitymagnetismdrivenantiferromagnetic}, which is in contrast to even-parity spin splitting observed in collinear altermagnets.
Compared to non-collinear magnetic configurations, collinear compensated magnets have larger range of available candidate materials, higher transition temperatures, and are much more accessible both theoretically and experimentally. 
Given the unique potential of odd-parity magnetism for spintronics, realizing such spin splitting in widely studied collinear antiferromagnets is highly desirable, particularly as they serve as promising platforms for next-generation high-temperature devices. 
Of equal importance is the realization of higher-order odd-parity spin splitting (such as $f$-wave or $h$-wave symmetry), which has remained largely unexplored. 

\begin{figure}[t]
\includegraphics[width=\linewidth]{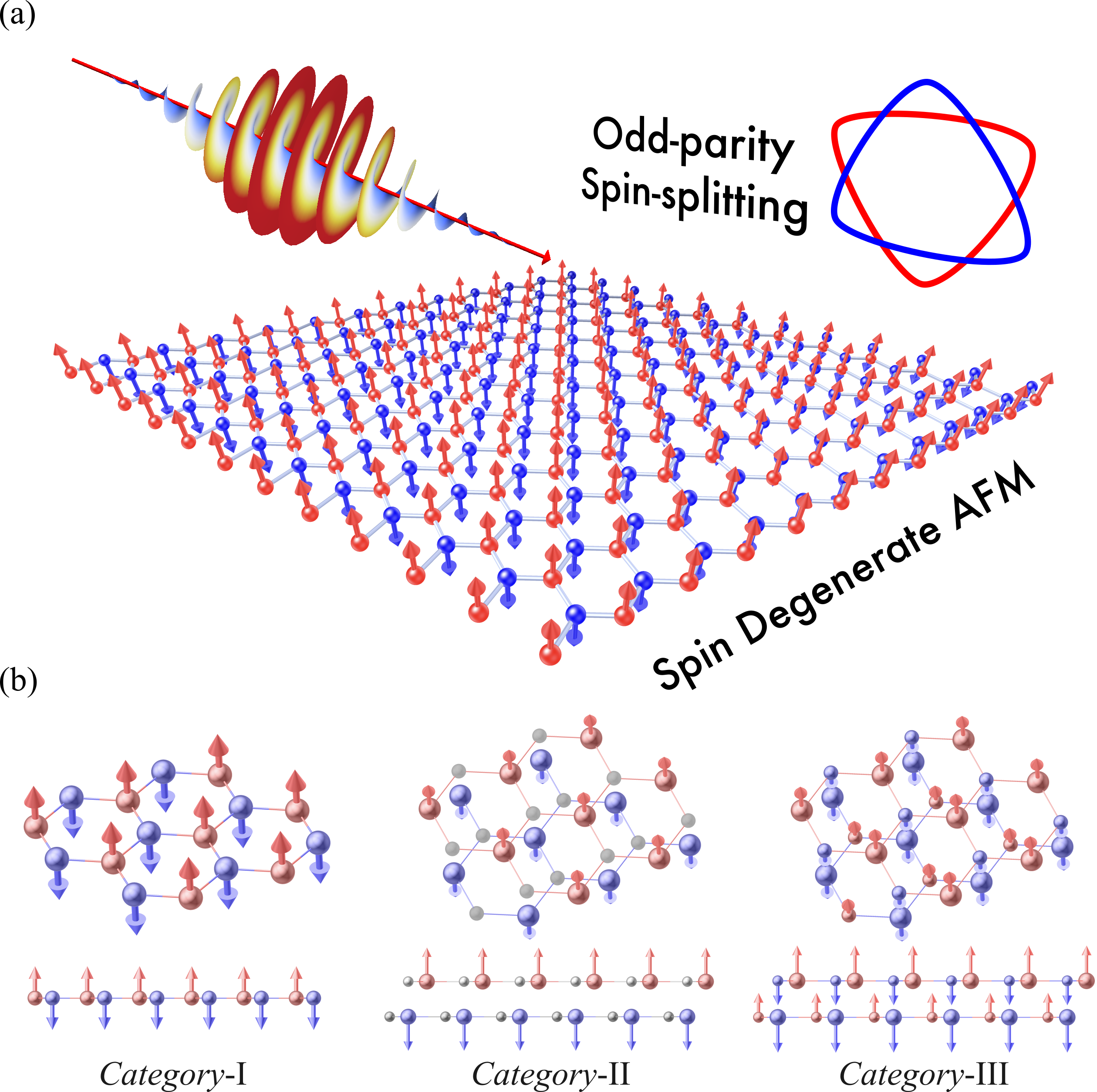}
\caption{\label{fig:fig1}
Illustration of a general route to the light-induced odd-parity spin splitting in conventional antiferromagnetism.
(a) The AFM system comprises two sublattices that are connected by $\left[ \mathcal{C}_{2}\bigparallel \mathcal{C}_{2z}   \right] $ or  $\left[ \mathcal{C}_{2}\bigparallel \mathcal{P}   \right] $ , schematically depicted by antiparallel red and blue spin arrows. 
The two sublattices form an inversion-symmetric pair, which gives rise to complete cancellation of spin splitting.
The incident light propagates along the $z$-axis with its polarization plane parallel to the $x$-$y$ plane.
The two sublattices exhibit opposite optical responses, leading to the emergence of $p$-wave and $f$-wave spin splitting in the bands of the AFM system  under circularly polarized light (CPL) irradiation.
(b) The schematic of three material candidate categories of $f$-wave spin splitting: Hexagonal lattice with Néel-type magnetic configuration (Category-I), AFM bilayers composed of FM monolayers. (Category-II), AFM bilayers composed of Ferrimagnetic monolayers (Category-III). 
}
\end{figure}

We note that Floquet engineering with periodic driving has recently emerged as a powerful approach for tuning symmetry-related electronic properties~\cite{2016PhysRevB.94.121106,2016PhysRevLett.117.087402,2016PhysRevB.94.081103,2017RevModPhys.89.011004,hubener2017creating,2018PhysRevLett.120.237403,2019PhysRevB.99.075121,2019PhysRevLett.123.206601,oka2019floquet,2020rudner2020band,2022bao2022light,2024zhan2024perspective,PhysRevB.110.L121118,2022PhysRevLett128066602,PhysRevLett.133.246606,ghorashi2025dynamical}, going beyond the well-established paradigm of static scenarios.  
Moreover, periodic light irradiation can  impose magnetic symmetries from the magnetic space group~\cite{2022PhysRevLett128066602}. Nevertheless, the prospect of dynamically controlling symmetry operators in the spin space group to achieve advanced manipulation of nonconventional magnetic properties, such as desirable spin splitting, is almost scarce. 
In this work, as schematically shown in Fig.~\ref{fig:fig1} (a),  we demonstrate that Floquet engineering can provide a reliable strategy to achieve odd-parity spin splitting in two-dimensional (2D) conventional antiferromagnetism under light irradiation. Based on symmetry arguments and effective model analysis, we show how to achieve $p$-wave and $f$-wave spin splitting from collinear antiferromagnetism. 
The symmetry requirements and distinct lattice models of potential candidates are established, see Fig.~\ref{fig:fig1} (b).
Remarkably, this light-induced odd-parity spin splitting exhibits flexible tunability through precise control of the crystalline symmetry or the polarization state of incident light. 
As concrete demonstrations, through first-principles calculations combining with Floquet theory, we verify the occurrence of $f$-wave spin splitting induced by circularly polarized light (CPL) in representative collinear antiferromagnets such as AFM $\mathrm{MnPS_3}$ monolayer, $\mathrm{FeCl_2}$ bilayer, and $\mathrm{NiRuCl_6}$ bilayer, and reversing the chirality of CPL makes a sign change of this odd-parity spin splitting in momentum space.  More results of other possible candidates are given in the Supplementary Material (SM)~\cite{SM}. Furthermore, we show that either symmetry breaking or irradiation with bicircular (BCL) or elliptically polarized light (EPL) can convert the light-induced spin polarization order from $f$-wave to $p$-wave.

\emph{\textcolor{blue}{Approach and symmetry analysis.}}---
For collinear AFM systems, the spin group is expressed as a direct product of a spin-only group ($\mathbf{r}_{s}$) and a nontrivial spin group ($\mathbf{R}_{s}$)~\cite{add1974,LB2022PhysRevX.12.031042,rn1l-d6cq}. Specifically, $\mathbf{r}_{s} $$=$$ \mathbf{C}_{\infty} + \bar{\mathcal{C}}_{2}\mathbf{C}_{\infty}$, where $\mathbf{C}_{\infty}$ represents spin rotations around the common axis of spins, and $\bar{\mathcal{C}}_{2} $$=$$ \mathcal{C}_2\mathcal{T}$, with $\mathcal{C}_2$ denoting a two-fold arbitrary spin rotation perpendicular to the axis of spins and $\mathcal{T}$ representing time reversal. Elements in $\mathbf{R}_{s}$ are of the form $[R_i || R_j]$, where $R_i \in \{E, \mathcal{C}_2\}$ acts in spin space ($E$ is the identity operation), and $R_j$ acts in crystal space~\cite{rn1l-d6cq}.
Thus, there are two kinds of symmetries that can   guarantee spin degeneracy: 
(I): $E\left[ \mathcal{C}_{2}\bigparallel \mathcal{O}' ,\mathcal{O}'\in\left\{ \tau,\mathcal{M}_{z} \right\}   \right] \epsilon_s\left(\boldsymbol{k}\right)$$=$$\epsilon_{-s}\left(\boldsymbol{k}\right)$; (II):
$\left[\bar{\mathcal{C}}_{2}\left[ \mathcal{C}_{2}\bigparallel \mathcal{O}'' ,\mathcal{O}''\in\left\{ \mathcal{P},\mathcal{C}_{2z} \right\}   \right]\right]^2$$=$$\left[\mathcal{O}''\mathcal{T}\right]^2$$=$$ -1$, indicating Kramers degeneracy. Here, $s = \pm 1$ denotes spin-up/down states at momentum $\bm{k}$, $\tau$ is fractional translation, $\mathcal{P}$ is spatial inversion, $\mathcal{M}_z$ is mirror reflection, and $\mathcal{C}_{2z}$ is two-fold rotation about $z$-axis.
For a CPL-driven collinear AFM system, the symmetry $\left[\mathcal{O}''\mathcal{T}\right]$ is broken by light filed. Consequently, 2D collinear AFM systems with $\left[ \mathcal{C}_{2}\bigparallel \mathcal{O}'\right]$  maintain spin degeneracy, whereas those with $\left[ \mathcal{C}_{2}\bigparallel \mathcal{O}''\right]$ exhibit spin splitting.
Remarkably, due to $\left[ \mathcal{C}_{2}\bigparallel \mathcal{O}''   \right] E_s\left(\boldsymbol{k}\right)=E_{-s}\left(\boldsymbol{-k}\right)$, the light-induced spin splitting is symmetry constrained to be odd-parity.
Naturally, the type of light-induced odd-parity spin splitting is further constrained by the symmetries of crystal or incident light. For example, $f$-wave ($p$-wave) spin splitting can be realized in system with  $\left[\mathcal{C}_2 \parallel \mathcal{C}_{6z}\right]$ or $\left[\mathcal{C}_2 \parallel \mathcal{S}_{6z}\right]$ ($\left[\mathcal{C}_2 \parallel \mathcal{C}_{2z}\right]$ or $\left[\mathcal{C}_2 \parallel \mathcal{P}\right]$) symmetry, as discussed in detail below and schematically shown in Fig.~\ref{fig:fig1} (a).

To clearly illustrate the light-induced spin splitting with odd-parity polarization in antiferromagnetism, we need to give a brief general description of the interaction between antiferromagnetism and periodic driving.
The incident light is described by a time-periodic gauge field $\mathbf{A}\left(t\right)=\mathbf{A}\left(t+T\right)$, where $T$=$2\pi/\omega$ denotes the optical period corresponding to the light frequency $\omega$.
The eigenstates of the light-driven system are periodic $\left|\psi \left(t\right) \right\rangle =\left|\psi \left(t+T\right) \right\rangle =e^{-i\epsilon t}\left|\varphi  \left(t\right) \right\rangle$, where $\left|\varphi \left(t\right)\right\rangle $ is dubbed Floquet states.
In the framework of Floquet theory~\cite{2017RevModPhys.89.011004,oka2019floquet,2020rudner2020band,2022bao2022light}, under the basis $\left|\varphi \left(t\right)\right\rangle =\sum_{n}e^{-in\omega}\left|u^{n}\right\rangle$, the time-dependent problem can be captured by Floquet Hamiltonian expressed in an infinite-dimensional extended Floquet space as
\begin{equation}
    \sum_{n}\left(H_{m-n}-n\omega\delta_{m,n}\right)\left|\varphi_{\alpha}^{n}\right\rangle =\epsilon_{\alpha}\left|\varphi_{\alpha}^{n}\right\rangle,
\end{equation}
where
$H_{m-n}\left(\boldsymbol{k}\right)=\frac{1}{T}\intop_{0}^{T}e^{i\left(m-n\right)\omega t}H\left(\boldsymbol{k},t\right)dt$, the time-dependent Hamiltonian $H\left(\boldsymbol{k},t\right)$ is  obtained by applying Peierls substitution to the original Hamiltonian under equilibrium, and integer $n$ is termed the Floquet index.
In the limit where the energy $\hbar\omega$ ($\hbar=1$ is adopted in this work) of the driving light is large compared to the other energy scales, the periodically driven system can be described by an effective Floquet Hamiltonian which reads
\begin{equation} \label{Eq2}
H_{\mathrm{eff}}\left(\boldsymbol{k}\right)=H_{0}\left(\boldsymbol{k}\right)+\frac{\left[H_{1}\left(\boldsymbol{k}\right),H_{-1}\left(\boldsymbol{k}\right)\right]}{\omega}+\mathcal{O}\left(\frac{1}{\omega^{2}}\right).
\end{equation}

\begin{figure}[t]
\includegraphics[width=\linewidth]{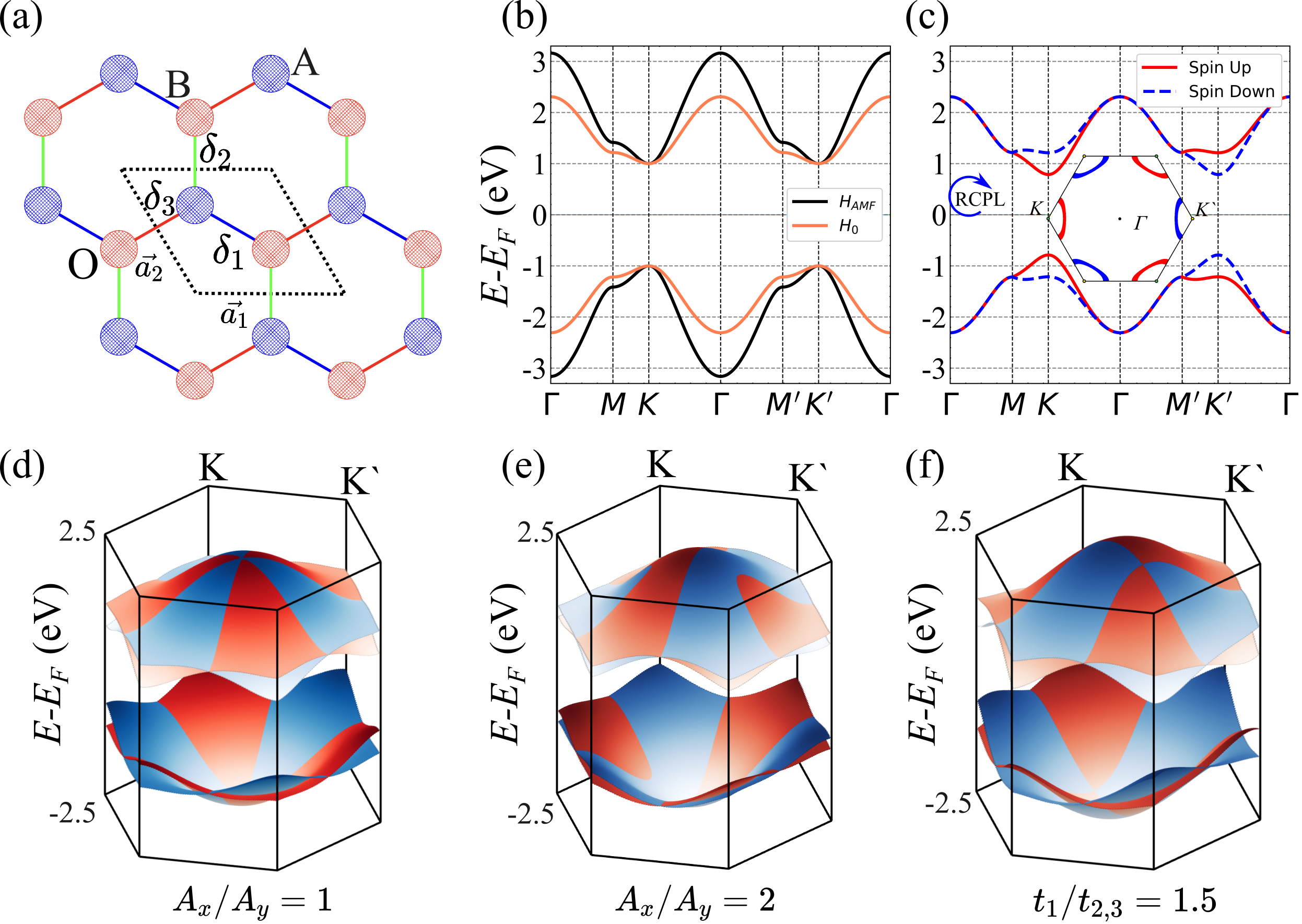}
\caption{\label{figure2}
A simple model for the light-induced odd-parity spin splitting and three categories of material candidates.
(a) The illustration of the hexagonal lattice model with  conventional AFM order.
In the schematic representation, spin-up and spin-down magnetic atoms are depicted as red and blue spheres, respectively.
(b) The spin-degenerate energy spectra of $H_{\mathrm{AFM}}\left(\boldsymbol{k}\right)$ and $H_{0}\left(\boldsymbol{k}\right)$.
(c) Spin-resolved energy spectra of conventional antiferromagnetism  under irradiation of right-handed CPL .
The corresponding spin-resolved isoenergy surfaces at $1$~eV that exhibit $f$-wave spin splitting are presented as inset in (c), and the corresponding three-dimensional band structures are shown in (d).
(e) Spin-resolved band structures with $p$-wave spin splitting in conventional antiferromagnetism irradiated by EPL with amplitude ratio $A_x/A_y$$=$$2$.
(f) CPL induced $p$-wave spin splitting on AFM hexagonal lattice with biaxial strain, i.e., $t_1/t_{2,3}$$=$$1.5$.
In panels (c)-(f), the parameters $t_{2,3}$$=$$m_1$$=$$-m_2$$=$$1$~eV and the light intensity of $e \mathrm{A}_0 / \hbar=e \mathrm{A}_y / \hbar = 0.5~\mathring{\mathrm{A}}^{-1}$ are adopted. Here, $t_i$ is the couplings along $\boldsymbol{\delta}_{i}$.
}
\end{figure}

Firstly, we show that linearly polarized light irradiation does not induce spin splitting in conventional AFM materials (see SM~\cite{SM} for a detailed discussion). 
Then, we turn to the periodic driving light fields such as circularly polarized light (CPL), elliptically polarized light (EPL), and bicircular light (BCL).
For CPL propagating along the $z$ direction, the electromagnetic gauge field is given by $\mathbf{A}\left(t\right)=\mathrm{A}_0\left(\eta \mathrm{sin} \omega t,\mathrm{cos} \omega t,0\right)$, with $\eta=+1$ ($\eta=-1$) representing the circular polarization of the left-(right-)hand.
For CPL irradiated systems with conventional AFM order, following the standard approach, one has
\begin{equation}
H_{0}\left(\boldsymbol{k}\right)=J_{0}\left(\mathrm{A}_{0}\left|\boldsymbol{\delta}\right|\right)\left[H_{\mathrm{AFM}}\left(\boldsymbol{k}\right)-H\left(\boldsymbol{m_{\zeta}}\right)\right]+H\left(\boldsymbol{m_{\zeta}}\right).
\end{equation}
Here, $H_{\mathrm{AFM}}\left(\boldsymbol{k}\right)$ is the Hamiltonian of the AFM system, $H\left(\boldsymbol{m_{\zeta}}\right)$ represents the exchange interaction originating from the AFM order with $\boldsymbol{m_{\zeta}}$ stands for the magnet moment at the site $\zeta$ ,
$J_0\left(x\right)$ is the zeroth-order Bessel function of $x$.
For simplicity, we consider only nearest-neighbor couplings and $\left|\boldsymbol{\delta}\right|$ denotes the distance between adjacent sites.
$H_{0}\left(\boldsymbol{k}\right)$  corresponds to scaling the coupling strength in $H_{\mathrm{AFM}}\left(\boldsymbol{k}\right)$ by a constant factor $J_{0}\left(\mathrm{A}_{0}\left|\boldsymbol{\delta}\right|\right)$, implying the absence of spin splitting in $H_{0}\left(\boldsymbol{k}\right)$, as shown in Fig.~\ref{figure2}(a) for the hexagonal AFM  lattice.
Thus, the potential light-induced spin splitting is completely determined by commutation $\left[H_{1}\left(\boldsymbol{k}\right),H_{-1}\left(\boldsymbol{k}\right)\right]=\omega H '\left(\boldsymbol{k}\right)$.

\emph{\textcolor{blue}{Model.}}---Here, we first demonstrate the emergence of $f$-wave spin splitting in the Floquet bands of a conventional antiferromagnet modeled on a hexagonal lattice composed of spinful $s$ orbitals (the corresponding wave functions are $\varphi_{\sigma,\{\sigma=\uparrow,\downarrow, X=A,B\}}^{X} $). 
This occurs under the preservation of either the $\left[\mathcal{C}_2 \parallel \mathcal{C}_{6z}\right]$ or $\left[\mathcal{C}_2 \parallel \mathcal{S}_{6z}\right]$ symmetry.
The lattice is illustrated in Fig.~\ref{figure2} (a), and the  AFM tight-binding Hamiltonian reads
\begin{eqnarray}\label{Eq4}
H & = & t_{NN}\sum_{\left\langle i,j\right\rangle }c_{i}^{\dagger}c_{j}+\sum_{i}c_{i}^{\dagger}\left[\mu_{\zeta}\sigma_{0}+\boldsymbol{m}_\zeta\cdot\boldsymbol{\sigma}\right]c_{i}+h.c.,
\end{eqnarray}
 where $i$ ($j$) labels sites in sublattice A (B) whose index is $\zeta$$=$$1$ ($\zeta$$=$$2$ ), $c_{i}^{\dagger}$$=$$\left[c_{i,\uparrow}^{\dagger},c_{i,\downarrow}^{\dagger}\right]$ ($c_{i}$$=$$\left[c_{i,\uparrow},c_{i,\downarrow}\right]$) is the two component creation (annihilation) operator of electron at site $i$.
The first term is the usual nearest-neighbor hopping term.
The AFM exchange
coupling and chemical potential are indicated by $\boldsymbol{m}_\zeta=m_\zeta\left(0,0,1\right)$ and $\mu_\zeta$, respectively.
Guaranteed by $\left[ \mathcal{C}_2\bigparallel \mathcal{C}_{6z}\right]$ ($\left[ \mathcal{C}_2\bigparallel \mathcal{S}_{6z}\right] $) symmetry, one has $m_1=-m_2$, and $\mu_1$$=$$\mu_2$$=$$0$.
Under the basis $\left[\varphi_{\uparrow}^{\mathrm{A}},\varphi_{\uparrow}^{\mathrm{B}},\varphi_{\downarrow}^{\mathrm{A}},\varphi_{\downarrow}^{\mathrm{B}}\right]$, by performing the Fourier transformation, the explicit form of the Hamiltonian Eq.~(\ref{Eq4}) can be written in momentum space as
\begin{eqnarray}
H_{\mathrm{AFM}}\left(\boldsymbol{k}\right) & = & \sigma_{0}\left[\begin{array}{cc}
0 & \Delta\left(\boldsymbol{k}\right)\\
\Delta\left(\boldsymbol{k}\right)^{*} & 0
\end{array}\right]\\ \nonumber
 &  & +\left(\left|m_{\zeta}\right|\sigma_{z}+\left|\mu_{\zeta}\right|\sigma_{0}\right)\tau_{z}
\end{eqnarray}
with $ \Delta{\left(\boldsymbol{k}\right)}=t_{NN}\sum_{i}e^{i\boldsymbol{k}\cdot\boldsymbol{\delta}_{i}}$, and $\boldsymbol{\delta}_{i} (i=1,2,3)$ are the nearest-neighbor vectors, as inserted in Fig.~\ref{figure2} (a), $\boldsymbol{\tau}$ and $\boldsymbol{\sigma}$ are orbital and spin Pauli matrices, respectively.

By a straightforward derivation, $H_{0}\left(\boldsymbol{k}\right)$ is given by $H_{\mathrm{AFM}}\left(\boldsymbol{k},t_{NN}\right) $$\mapsto$$ H_{\mathrm{AFM}}\left(\boldsymbol{k},t_{NN}J_0(\frac{aA_0}{\sqrt{3}})\right)$, with $a=1$ is  lattice constant.
The spin-degenerate energy spectra of $H_{\mathrm{AFM}}\left(\boldsymbol{k},t_{NN}\right)$ and $H_{0}\left(\boldsymbol{k}\right)$ are given in Fig.~\ref{figure2} (b), where no spin splitting is found.
Thus, the second term in Eq.~(\ref{Eq2})  determines the potential spin splitting.
And, this term is expressed as
\begin{eqnarray}\label{Eq6}
\omega H'\left(\boldsymbol{k}\right) & = & 8\sqrt{3}\eta J_{1}\left(\frac{A_{0}}{\sqrt{3}}\right)^{2}t_{NN}^{2}\prod_{i}\mathrm{sin}\left(k'_{i}/2\right)\sigma_{0}\tau_{z}\nonumber \\
 & = & f\left(\boldsymbol{k}\right)\sigma_{0}\tau_{z}
\end{eqnarray}
with $\left[ k'_{1},k'_{2},k'_{3}\right]$=$\left[  k_{1},k_{2},k_{1}+k_{2}\right]$ where $\boldsymbol{k}$$=$$k_1 \boldsymbol{b}_1+k_2 \boldsymbol{b}_2$ with $\boldsymbol{b}_{1,2}$ being reciprocal lattice vectors.
Thus, the eigenvalues for the light-driven
AFM system are obtained as
\begin{equation}\label{Eq7}
E_{s}\left(\boldsymbol{k}\right)=\pm\sqrt{\left|J_0(\frac{A_0}{\sqrt{3}})\Delta\left(\boldsymbol{k}\right)\right|^{2}+\left(\left|m_{\zeta}\right|+sf\left(\boldsymbol{k}\right)/\omega\right)^{2}}.
\end{equation}
From Eq.~(\ref{Eq7}), one finds that the light-irradiated AFM system and $E_{s}\left(\boldsymbol{k}\right)$  exhibit four key features:

\noindent I: Due to $H'\left(k'_i=0\right)\equiv\boldsymbol{0}$, the spin degeneracy is maintained along high-symmetry lines $k'_i=0$.

\noindent II: $E_{s}\left(\boldsymbol{k}\right)$=$E_{-s}\left(-\boldsymbol{k}\right)$, as proved in the last section and shown in Fig.~\ref{figure2} (c).

\noindent III: The spin splitting adheres to the crystal symmetry $[E || \mathcal{C}_{3z}]$, i.e.,  $E_{s}\left(\boldsymbol{k}\right)=E_{s}\left(\mathcal{C}_{3z}^{-1}\boldsymbol{k}\right)$. Combining with features I$\sim$II, as inserted in Figs.~\ref{figure2} (c) and \ref{figure2} (d), we obtain $f$-wave spin splitting.

\noindent IV: The appearance of $\tau_z$ in Eq.~(\ref{Eq6}) implies that spin splitting also can be regarded as the opposite response of the two sublattices to CPL.

To bridge the gap between theoretical modeling and material prediction, 
following our symmetry analysis,
we identify three distinct categories of candidate materials (designated \textit{Category}-I through  \textit{Category}-III) for realizing $f$-wave spin splitting in AFM system, as schematically illustrated in Fig.~\ref{fig:fig1} (b).
\textit{Category}-I: As detailed discussed, CPL enables $f$-wave spin splitting in a hexagonal monolayer with Néel-type magnetic configuration.
When such monolayers are stacked in a van der Waals magnetic multilayer, $f$-wave spin splitting can also be achieved by CPL (see SM~\cite{SM} for details).
\textit{Category}-II:
The light induced $f$-wave spin splitting can also be achieved in AFM bilayers composed of FM monolayers.
\textit{Category}-III: One can also replace the monolayers employed in \textit{Category}-II by ferrimagnetic (even fully compensated ferrimagnetic) monolayer.
In SM~\cite{SM}, the light-induced $f$-wave spin splitting in lattices belong to \textit{Category}-II and \textit{Category}-III are discussed.

In addition to generating $f$-wave spin splitting, we show that the light-induced spin splitting can be effectively modulated through the polarization of incident light and the engineering of crystalline symmetries, thereby enabling the realization of $p$-wave spin splitting.
As directly inferable from Eq.~(\ref{Eq4}),  turning the helicity of the incident circularly polarized light from right-handed ($\eta $=$ 1$) to left-handed ($\eta $=$ -1$) leads to an reversal of the spin-splitting~\cite{SM}.
Furthermore, modifying the symmetry of incident light can give rise to $p$-wave spin splitting~\cite{SM}.
To this end, we employ BCL and EPL propagating along the $z$-direction.
Their vector potentials are expressed respectively as $\mathbf{A}_{\mathrm{EPL}}\left(t\right)=\left(\mathrm{A}_x \mathrm{sin} \omega t,\mathrm{A}_y\mathrm{cos} \omega t,0\right)$ and $\mathbf{A}_{\mathrm{BCL}}\left(t\right)$$=$$\sqrt{2}\mathrm{A}_{0}\mathrm{Re}\left[e^{-i\omega_{1}t},e^{-i\kappa \omega_{1}t},0\right]$ with $\kappa$ denoting the frequency ratio between the right- and left-handed components constituting the BCL.
In this case, although analytically deriving the explicit form of  $H'\left(\boldsymbol{k}\right)$ and corresponding eigenvalues is highly cumbersome, the numerical results can be readily obtained via Eq.~(\ref{Eq2}) (see more details in \cite{SM}).
Fig.~\ref{figure2}(e) shows that, in the presence of elliptical polarized light (EPL) with an amplitude ratio of $A_x/A_y = 2$, the energy spectrum of AFM system exhibits $p$-wave spin splitting.

We now investigate the $p$-wave spin splitting induced by CPL in conventional antiferromagnetism. 
To achieve distinct spin polarization orders, we begin by reducing the spatial symmetries of the lattice model described in Eq.~(\ref{Eq4}).
Applying uniaxial strain along the [100] and [110] directions lowers the point group symmetry of the hexagonal lattice shown in Fig.~\ref{figure2}(a) from $D_{6h}$ to $C_{2h}$ and $D_{2h}$, respectively.
As illustrated in Fig.~\ref{figure2}(f), taking the strain along [110] as an example, the energy bands exhibit $p$-wave spin polarization order under CPL irradiation.

\begin{figure}
\includegraphics[width=\linewidth]{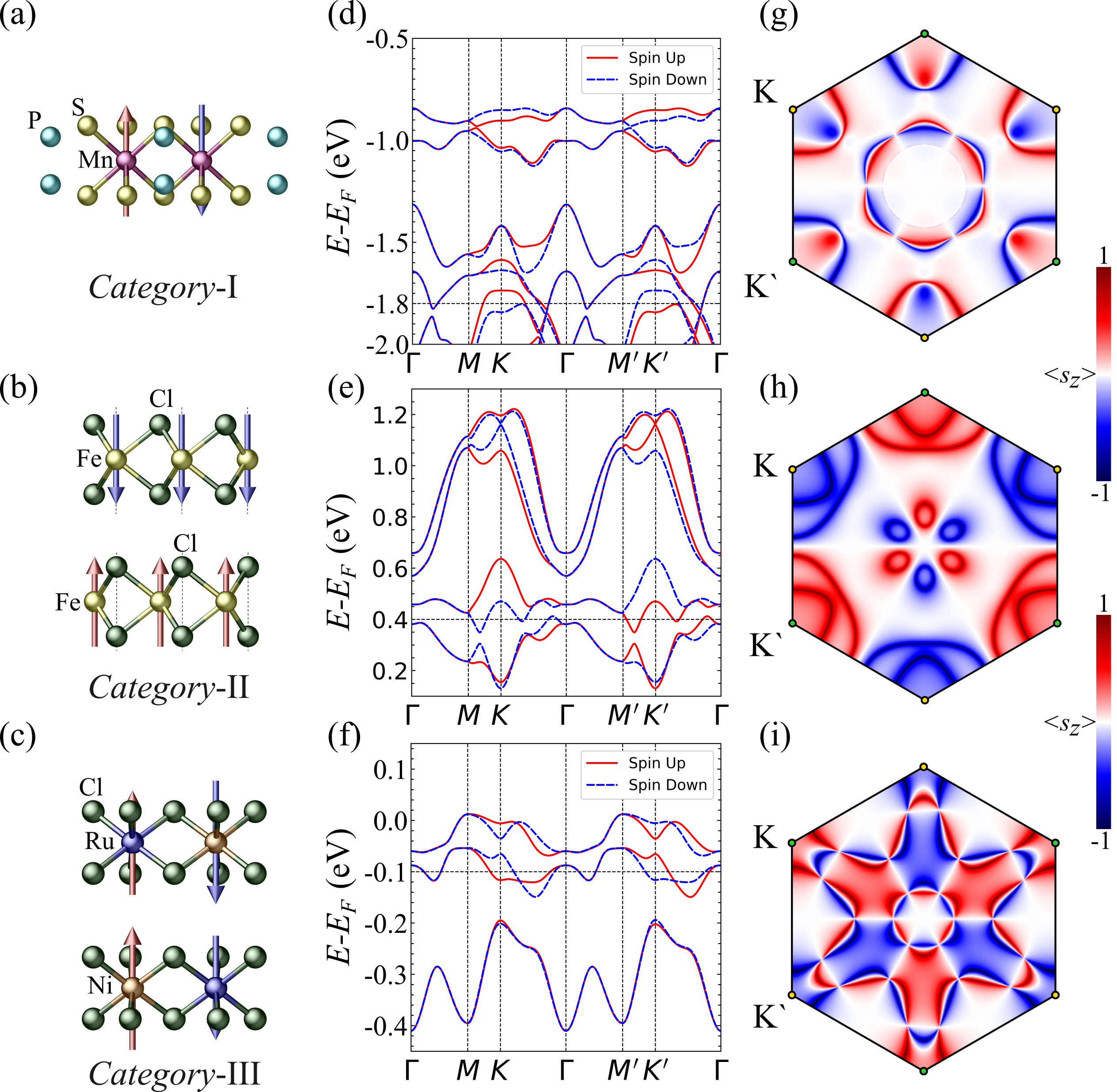}
\caption{\label{fig:fig3}CPL induced $f$-wave spin splitting in representative materials, i.e., AFM $\mathrm{MnPS_3}$ monolayer (\textit{Category}-I),  AFM $\mathrm{FeCl_2}$ bilayer (\textit{Category}-II), and   AFM $\mathrm{NiRuCl_6}$ bilayer (\textit{Category}-III).
(a)-(c) The side-view and bird-view of crystal structure and magnetic order of  AFM $\mathrm{MnPS_3}$ monolayer,  AFM $\mathrm{FeCl_2}$ bilayer, and   AFM $\mathrm{NiRuCl_6}$ bilayer.
(d)-(f) The spin-resolved band structures (curves in red solid represent the spin-up state ) along high-symmetry lines of the  AFM $\mathrm{MnPS_3}$ monolayer,  AFM $\mathrm{FeCl_2}$ bilayer, and   AFM $\mathrm{NiRuCl_6}$ bilayer under irradiation of CPL with a light intensity of $e \mathrm{A}_0 / \hbar = 0.3~\mathring{\mathrm{A}}^{-1}$ and the photon energy of this light $\hbar \omega= 10$~$\text{eV}$.
The light irradiation  preserves the spin degeneracy along the high-symmetry line $\Gamma$-M while lifting the degeneracy at  at other arbitrary $\boldsymbol{k}$-point.
(g)-(i) The spin-resolved isoenergy surfaces with $f$-wave spin splitting of the AFM $\mathrm{MnPS_3}$ monolayer at $-1.8$~eV, the AFM $\mathrm{FeCl_2}$ bilayer at $0.4$~eV, and the AFM $\mathrm{NiRuCl_6}$ bilayer at $-0.1$~eV.
}
\end{figure}

\emph{\textcolor{blue}{Material realization of odd-parity spin splitting.}}---
With the theoretical framework in place, we now shift focus from theoretical design concepts to the material realization of the light-induced odd-parity spin splitting in conventional antiferromagnetism.
According to the symmetry analysis established above, 
the AFM system should be constrained by $\left[ \mathcal{C}_{2}\bigparallel \mathcal{O}''\right]$  rather than $\left[ \mathcal{C}_{2}\bigparallel \mathcal{O}'\right]$.
Here, by density functional theory calculations, we predict three candidate materials (see Figs. \ref{fig:fig3}(a)-\ref{fig:fig3}(c)) for realizing $f$-wave spin splitting: monolayer $\mathrm{MnPS_3}$, bilayer $\mathrm{FeCl_2}$, and bilayer $\mathrm{NiRuCl_6}$, corresponding to \textit{Category}-I, -II, and -III, respectively.
Notably, all three candidate materials have been extensively studied, with experimentally confirmed ground-state magnetic structures consistent with the required symmetry constraints.

Through first-principles calculations (see the SM~\cite{SM} for calculation details), we first confirm the intrinsic spin-degenerate band structure in the three types of candidates. 
Then we apply a counterclockwise CPL 
to $\mathrm{MnPS_3}$, $\mathrm{FeCl_2}$ and $\mathrm{NiRuCl_6}$.
The photon energy of this light is set as $\hbar \omega$=$ 10$~$\text{eV}$ to meet the high-frequency approximation condition.
Figures.~\ref{fig:fig3}(d)-\ref{fig:fig3}(f) presents the calculated spin-resolved band structures of $\mathrm{MnPS_3}$, $\mathrm{FeCl_2}$ and $\mathrm{NiRuCl_6}$  under the light intensity of $e \mathrm{A}_0/\hbar = 0.3 ~\mathring{\mathrm{A}}^{-1} $ .
The calculated results show that light irradiation lifts the spin degeneracy at generic $\boldsymbol{k}$-points while preserving it along the high-symmetry $\Gamma$–M line.
The spin-resolved isoenergy surfaces for $\mathrm{MnPS_3}$ at -1.8~eV, $\mathrm{FeCl_2}$ at 0.4~eV and $\mathrm{NiRuCl_6}$ at -0.1~eV are presented in Figs.~\ref{fig:fig3}(g)-\ref{fig:fig3}(i),  confirming the appearance of light-induced $f$-wave spin splitting in these three classes of AFM system.

Before concluding, using the AFM $\mathrm{FeCl_2}$ bilayer as an example, we demonstrate the tunability of $f$-wave spin polarization order via precise control of the light polarization and crystalline symmetry (see SM~\cite{SM} for details).
We apply counterclockwise BCL, EPL, uniaxial strain along the [110] or [100] direction to manipulate the light-induced $f$-wave spin polarization order in AFM $\mathrm{FeCl_2}$ bilayer system. 
Owing to the breaking of the $\left[\mathcal{C}_2 \parallel \mathcal{S}_{6z}\right]$, the application of BCL, EPL, or strain combined with CPL leads to $p$-wave spin splitting, as shown in Fig.S7 and Fig.S8 ~\cite{SM}. These observations are fully consistent with the theoretical analysis presented in previous sections.

\emph{\textcolor{blue}{Conclusions and discussion.}}---Summarily, we present a universal approach for achieving and controlling odd-parity spin splitting in conventional antiferromagnetism through irradiation of light.
Employing low-energy effective models alongside symmetry analysis, we point out the symmetry requirements for achieving light-induced $f$- and $p$-wave spin splitting.
Three distinct lattice categories are predicted to exhibit $f$-wave spin splitting.
Due to the constraint from crystalline symmeties, the $h$- wave or higher-order light-induced spin splitting may be found in 2D quasicrystal systems.
Using first-principles calculations, we identify the extensively studied AFM $\mathrm{MnPS_3}$ monolayer, AFM $\mathrm{FeCl_2}$ bilayer, and AFM $\mathrm{NiRuCl_6}$ bilayer as promising candidates for the realization of $f$-wave spin splitting via CPL irradiation.
Notably, beyond the materials discussed in the main text, compounds such as MnBi$_2$Te$_4$, RuO$_2$ (1H phase) and $\mathrm{MnX_2}$ ($X$$=$ S, Se, Te) (see Table SI of the SM~\cite{SM} for additional candidates) also provide promising platforms for realizing of odd-parity spin splitting, which may be detected experimentally with time-resolved and angle-resolved photoemission spectroscopy (TrARPES) and time-resolved transport measurements \cite{RevModPhys.93.041002,RevModPhys.96.015003,SM}.
Moreover, compared with non-collinear systems, the higher Néel temperatures of these collinear antiferromagnets provide critical thermal stability under optical driving, ensuring the feasibility of the experiment.
On the other hand, even if the photon energy is much smaller than the bandwidth, light-induced odd-parity spin splitting is expected to occur, making the experimental attempt more practical, see DFT results with $\hbar\omega=0.1$~eV and light intensity of $e \mathrm{A}_0 / \hbar = 0.2~\mathring{\mathrm{A}}^{-1}$ in SM~\cite{SM}.
Furthermore, the odd-parity spin splitting is verified to be flexibly controlled by manipulating the polarization of the incident light and crystalline symmetry, enabling reversal or conversion of spin-splitting. 
In light of the recent experimental realization of non-collinear $p$-wave magnets, our proposed approach for achieving and manipulating collinear $p$-wave and $f$-wave magnets paves the way for discovering promising candidates for next-generation spintronic devices, including high-density magnetic memories and terahertz nano-oscillators.

\emph{\textcolor{blue}{Note added.}}---Recently, we have noticed some independent research findings that overlap with our results
\cite{li2025floquet,zhu2025floquet}.

\emph{\textcolor{blue}{Acknowledgments.}}---This work was supported by the National Natural Science Foundation of China (NSFC, Grants No. 12222402, No. 92365101, No. 12347101, and No.~12204074), the Natural Science Foundation of Chongqing (Grants No. 2023NSCQ-JQX0024 and No. CSTB2022NSCQMSX0568), and Beijing National Laboratory for Condensed Matter Physics (Grant No. 2024BNLCMPKF015). 
	
\bibliography{Manuscript}

\clearpage
\onecolumngrid
\begin{center}
{\bf Supplemental materials of ``Light-induced Odd-parity Magnetism in Conventional Collinear Antiferromagnets"}
\end{center}
\title{Supplemental materials of ``Light-induced Odd-parity Magnetism in Conventional  Antiferromagnetism"}

This supplementary material is organized in eight sections. 
In the first section, we provide the details of DFT calculations in the main text. 
In the second section, we clarify the spin degeneracy in light-irradiated conventional AFM with $\left[ \mathcal{C}_{2}\bigparallel \mathcal{O}' ,\mathcal{O}'\in\left\{ \tau,\mathcal{M}_{z} \right\}   \right] $. 
In the third section, we give more in-depth discussion on light irradiated hexagonal lattices with tight-binding models. 
In the fourth section, we show the intrinsic band structures of materials in the main text. 
In the fifth section, we discussed the spin-splitting in BCL irradiated conventional antiferromagnetism.
In the sixth section, we show the $p$-wave spin splitting in $\mathrm{FeCl_2}$ induced by BCL, EPL, and strain.
In the seventh section, we provide the list of material candidates for achieving odd-parity spin splitting.
 The last section provides a brief discussion on experimental realization of light-induced odd parity spin splitting.


\section{DFT calculations}
We employed density functional theory (DFT)~\cite{hohenberg1964density,kohn1965self} within the Vienna Ab initio Simulation Package (VASP) using the projector augmented-wave method~\cite{blochl1994projector}. The generalized gradient approximation (GGA)~\cite{perdew1996generalized} of Perdew-Burke-Ernzerhof type was used to account for the exchange-correction potential. The kinetic-energy cutoff was set to 500~eV, and a $18\times18\times1$ $\Gamma$-centered $k$-point mesh was used for all calculations. A criterion of energy different in electronic self-consistent calculation is set to $10^{-8}$~eV. The lattice consists are fully relaxed and converted below $10^{-3}$~eV/{\AA} for residual force.

To utilize Floquet laser light field in our material system, we first projected the plane waves basis for first-principle calculations to  maximally localized Wannier functions (MLWF) basis, with the help of Wannier90~\cite{mostofi2008wannier90,wuWannierToolsOpensourceSoftware2018}.

The Wannier tight-binding Hamiltonian we gain can be expressed as
\begin{eqnarray}
H(\textbf{r} )=\sum_{mn}^{}  \sum_{j} t_{j}^{mn} C^{\dagger } _{m}(\textbf{R}_{j})C _{n}(\textbf{R}_{0})+h.c.
\end{eqnarray}
where $t^{mn}_{j}$ is the hopping amplitude from $m$ localized Wannier orbital of site $R_0$ to $n$ localized Wannier orbital of site $R_j$. The operator $C_m^{\dagger}(R_j)$ ($C_m(R_0)$) create and (annihilates) a electron on respected sites. The vector potential $\textbf{A}(t)=\textbf{A}(t+T)$ of time-periodic CPL coupled to the Hamiltonian by Peierls substitution as. 
\begin{eqnarray}
t_{j}^{mn}(t)=t_j^{mn}e^{i\frac{e}{\hbar}\textbf{A}(t)\cdot \textbf{d}_{mn}}
\end{eqnarray}
Where $\textbf{d}_{mn}$ is the relative position vector between two Wannier orbitals. Thus we obtained the time-dependent Hamiltonian as 
\begin{eqnarray}
    H(k,t)=\sum_{mn}\sum _{j}t_j^{mn}(t)e^{ik\cdot \textbf{R}_j} C^{\dagger }_m(k,t)C_n(k,t )+h.c.
\end{eqnarray}
Due to the translation symmetry of both the time and lattice, Floquet theorem can be effectively applied to the time-dependent Hamiltonian. Thus we express the Floquet-Bloch creation (annihilation) operator as 
\begin{eqnarray}
C_m^{\dagger }(k,t )=\sum_j\sum^{\infty }_{\alpha =-\infty }C_{\alpha m}^{\dagger }(\textbf{R}_j )e^{+ik\cdot \textbf{R}_j-i\alpha \omega t  } 
\end{eqnarray}
\begin{eqnarray}
C_m(k,t )=\sum_j\sum^{\infty }_{\alpha =-\infty }C_{\alpha m}(\textbf{R}_j )e^{-ik\cdot \textbf{R}_j+i\alpha \omega t  } 
\end{eqnarray}
Thus, we gain an effective static Hamiltonian in frequency an momentum space 
\begin{eqnarray}
    H(k,\omega  )=\sum_{m,n} \sum_{\alpha ,\beta }[H^{mn}_{\alpha - \beta }+H_{\Omega}]C^{\dagger }_{\alpha m}(k )C_{\beta n}(k )+h.c.
\end{eqnarray}
with
\begin{eqnarray}\label{Eq7}
    H^{mn}_{\alpha -\beta }=\sum_j e^{ik\cdot \textbf{R}_j  }\frac{1}{T} \int_{0}^{T}t_{j}^{mn}e^{i\frac{e}{\hbar } \textbf{A}(t)\cdot \textbf{d}_{mn}}e^{i(\alpha -\beta )\omega t} dt  
\end{eqnarray}
\begin{eqnarray}
    H_{\Omega}=\alpha \hbar \omega \delta_{mn}\delta_{\alpha \beta }
\end{eqnarray}
where $\hbar\omega$ represent the photon energy, $(\alpha,\beta)$ is the Floquet index ranging from $-\infty $ to $\infty$ . Applying Floquet theory in the high-frequency limit, the periodically driven system can be described by the effective Floquet-Bloch Hamiltonian\cite{PhysRevB.25.6622,PhysRevA.38.1739,PhysRevA.68.013820,PhysRevLett.91.110404,PhysRevX.4.031027}, which can be described as
\begin{equation} \label{Eq9}
H_{\mathrm{eff}}\left(\boldsymbol{k},\omega\right)=H_{0}\left(\boldsymbol{k},\omega\right)+\frac{\left[H_{1}\left(\boldsymbol{k},\omega\right),H_{-1}\left(\boldsymbol{k},\omega\right)\right]}{\omega}+\mathcal{O}\left(\frac{1}{\omega^{2}}\right).
\end{equation}


\section{Spin degeneracy in light-irradiated conventional AFM with $\left[ \mathcal{C}_{2}\bigparallel \mathcal{O}' ,\mathcal{O}'\in\left\{ \tau,\mathcal{M}_{z} \right\}   \right] $}
\begin{figure}[h]
\includegraphics[width=0.75\linewidth]{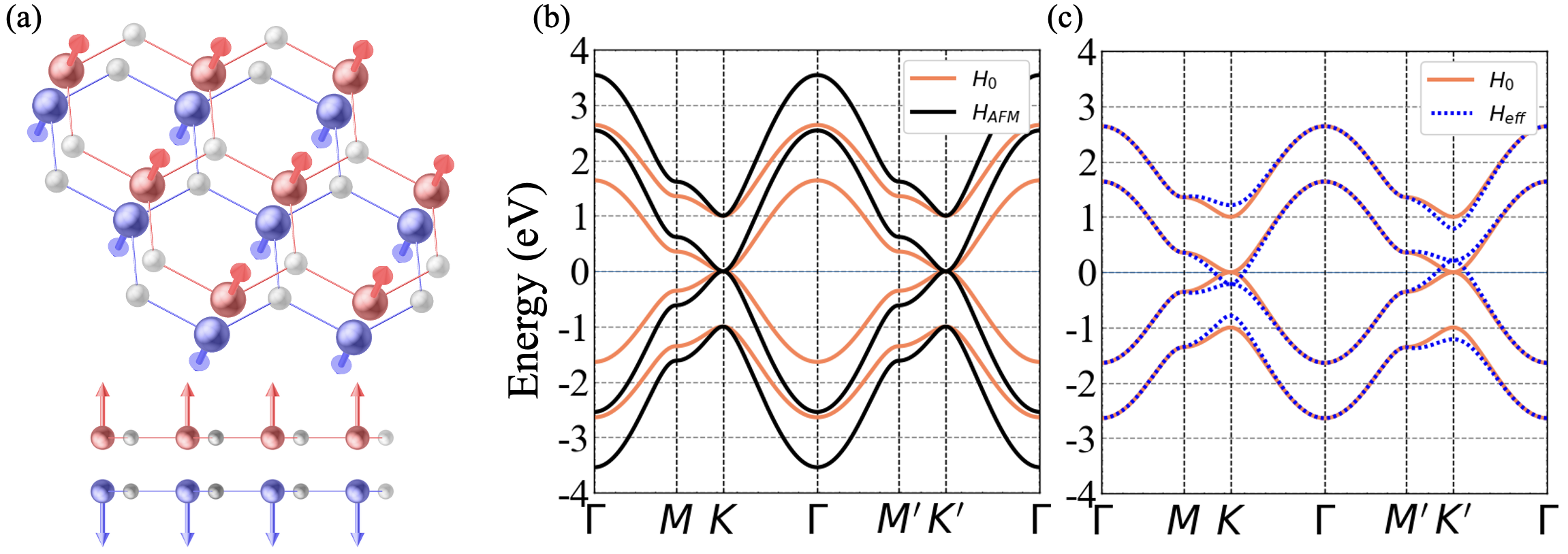}
\caption{\label{fig:sq}
CPL-irradiated conventional AFM with $\left[ \mathcal{C}_{2}\bigparallel \mathcal{O}' ,\mathcal{O}'\in\left\{ \tau,\mathcal{M}_{z} \right\}   \right] $.
(a) The illustration figure of AFM bilayer honeycomb lattice with $\left[ \mathcal{C}_{2}\bigparallel \mathcal{M}_{z}  \right] $.
(b)-(c) The spin-degenerate energy spectra of $H_{\mathrm{AFM}}\left(\boldsymbol{k}\right)$ and $H_{0}\left(\boldsymbol{k}\right)$ and $H_{\mathrm{eff}}\left(\boldsymbol{k}\right)$. The parameters of nearest neighbor hopping $t=1$~eV and magnetic moment $m=1$~eV and the light intensity of $e \mathrm{A}_0 / \hbar = 0.5~\mathring{\mathrm{A}}^{-1}$ are adopted.
}
\end{figure}

In this section, we demonstrate that circularly polarized light cannot induce spin splitting in antiferromagnetic systems with  $\left[ \mathcal{C}_{2}\bigparallel \mathcal{O}' ,\mathcal{O}'\in\left\{ \tau,\mathcal{M}_{z} \right\}   \right] $ symmetry.
We begin with a tight-binding model on an AFM hexagonal lattice belonging to space group $P\overline{6}$ (No. 174), as illustrated in Fig.~\ref{fig:sq}. Magnetic atoms occupy the Wyckoff position $2h$ at $(1/3, 2/3, z)$, while additional nonmagnetic atoms reside at the $2i$ position: $(2/3, 1/3, z)$.
The spin degeneracy in such conventional antiferromagnets is protected by the spin-space group symmetry $\left[ \mathcal{C}_{2}\bigparallel \mathcal{M}_{z}  \right] $. Two key points are noteworthy:
(i) the mirror symmetry $\mathcal{M}_{z}$ maps $(x, y, z)$ to $(x, y, -z)$;
(ii) the $z$-component of the time-periodic gauge field $\mathbf{A}(t)$ vanishes.
As a result, the two magnetic sublattices with opposite magnetic moment respond identically under light irradiation, thereby precluding spin splitting.
Figs.~\ref{fig:sq}(b)-(c) demonstrate that, even $H'\left(\boldsymbol{k}\right)$ is non-zero, the energy bands of $H_{\mathrm{AFM}}\left(\boldsymbol{k}\right)$, $H_{0}\left(\boldsymbol{k}\right)$, and $H_{\mathrm{eff}}\left(\boldsymbol{k}\right)$ exhibit band-degenerate characteristics.
Similarly, in case of antiferromagnetic systems with  $\left[ \mathcal{C}_{2}\bigparallel  \tau   \right] $ symmetry, there is no light-induced spin splitting.

\section{Spin degeneracy and spin splitting in light irradiated hexagonal lattices}
In this section, we give more detailed discussion on the honeycomb lattice tight-binding model in the third section of the main text. 
The model in the main text is
\begin{eqnarray}\label{Eq10}
H & = & t\sum_{\left\langle i,j\right\rangle }c_{i}^{\dagger}c_{j}+\sum_{i}c_{i}^{\dagger}\left[\mu_{\zeta}\sigma_{0}+\boldsymbol{m}_{\zeta}\cdot\boldsymbol{\sigma}\right]c_{i}+h.c.,
\end{eqnarray}
which can be expressed in momentum space as, under the basis $\left[\varphi_{\uparrow}^{\mathrm{A}},\varphi_{\uparrow}^{\mathrm{B}},\varphi_{\downarrow}^{\mathrm{A}},\varphi_{\downarrow}^{\mathrm{B}}\right]$, 
\begin{eqnarray}\label{Eq11}
H_{\mathrm{AFM}}\left(\boldsymbol{k}\right) & = & \sigma_{0} \left[\begin{array}{cc}
0 & \Delta\left(\boldsymbol{k}\right)\\
\Delta\left(\boldsymbol{k}\right)^{*} & 0
\end{array}\right]\\ \nonumber
 &  & +\left(\left|m_{\zeta}\right|\sigma_{z}+\left|\mu_{\zeta}\right|\sigma_{0}\right)\tau_{z}
\end{eqnarray}
with $ \Delta{\left(\boldsymbol{k}\right)}=t\sum_{i}e^{i\boldsymbol{k}\cdot\boldsymbol{\delta}_{i}}$, and $\boldsymbol{\delta}_{i} (i=1,2,3)$ are the nearest-neighbor vectors.

\subsection{linear polarized light irradiated AFM honeycomb lattice}

\begin{figure}[h]
\includegraphics[width=0.35\linewidth]{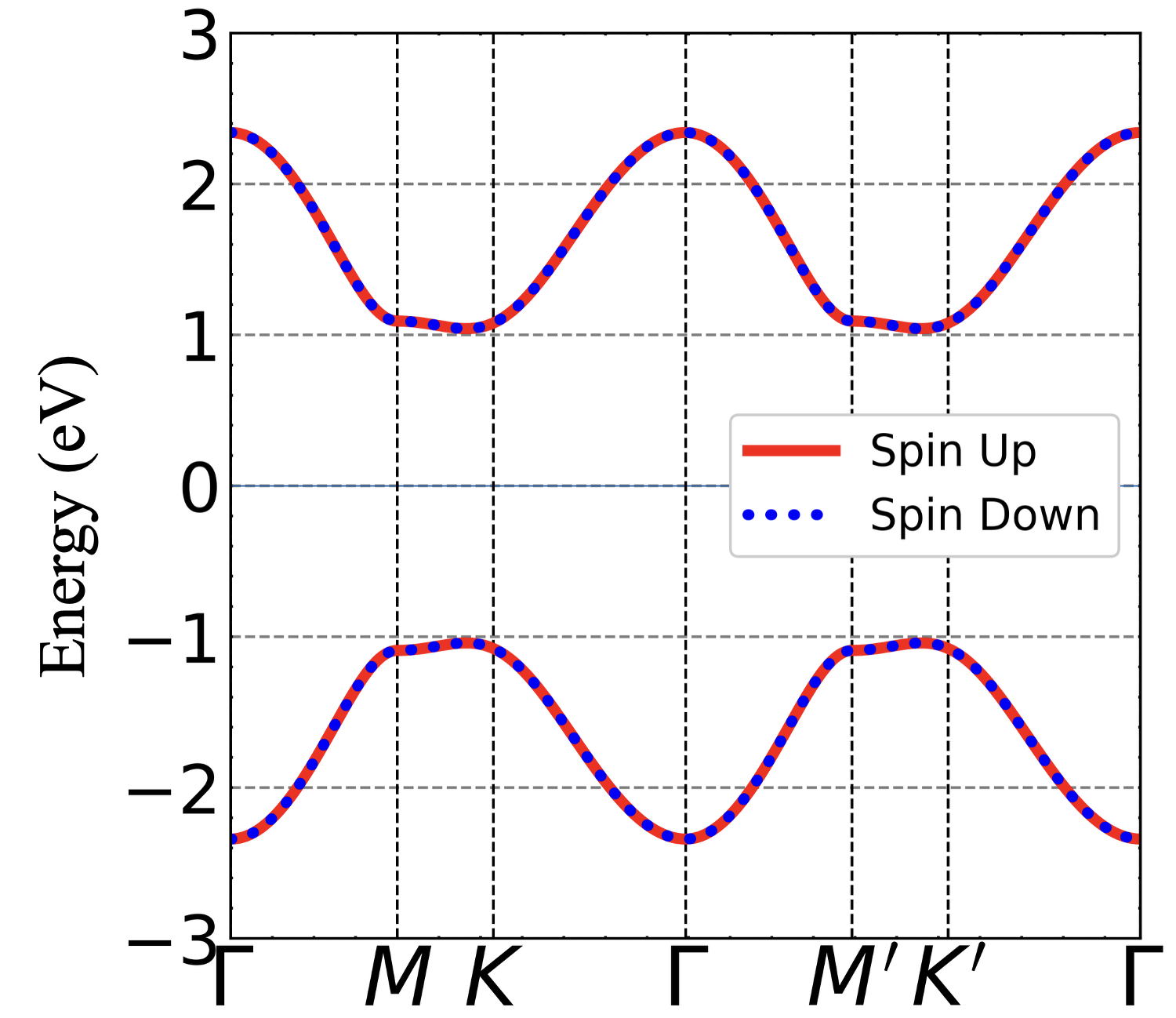}
\caption{\label{fig:LPL}The band structure of the lattice model in Eq.2 of the main text under LPL with a vector potential $\mathbf{A}_{\mathrm{LPL}}\left(t\right)=\left(\mathrm{A}_x \mathrm{sin} \omega t,0,0\right)$.
The parameters $t=m_{\zeta}=1$~eV, $\mu_{\zeta}=0$, and the light intensity of $e \mathrm{A}_0 / \hbar = 0.5~\mathring{\mathrm{A}}^{-1}$ are adopted.
}
\end{figure}

First, we show that spin-splitting can not be induced by linear polarized light (LPL). We apply an LPL irradiated along the $z$ direction with $x$-axis polarization to AFM honeycomb lattice model Eq. (\ref{Eq11}).
With straightforward derivation, combining Eq. (\ref{Eq7}) and Eq. (\ref{Eq9}), one has $H_{1}\left(\boldsymbol{k},\omega\right)=-H_{-1}\left(\boldsymbol{k},\omega\right)$, which results in $\left[H_{1}\left(\boldsymbol{k}\right),H_{-1}\left(\boldsymbol{k}\right)\right]=\omega H '\left(\boldsymbol{k}\right)=0$.
Thus, as shown in Fig.~\ref{fig:LPL},  with the light intensity of $e \mathrm{A}_0 / \hbar = 0.5~\mathring{\mathrm{A}}^{-1}$ as an example, the band structure of the LPL irradiated AFM honeycomb lattice is spin-degenerate.

\subsection{CPL irradiated AFM honeycomb lattice with next nearest couplings}

\begin{figure}
\includegraphics[width=0.5\linewidth]{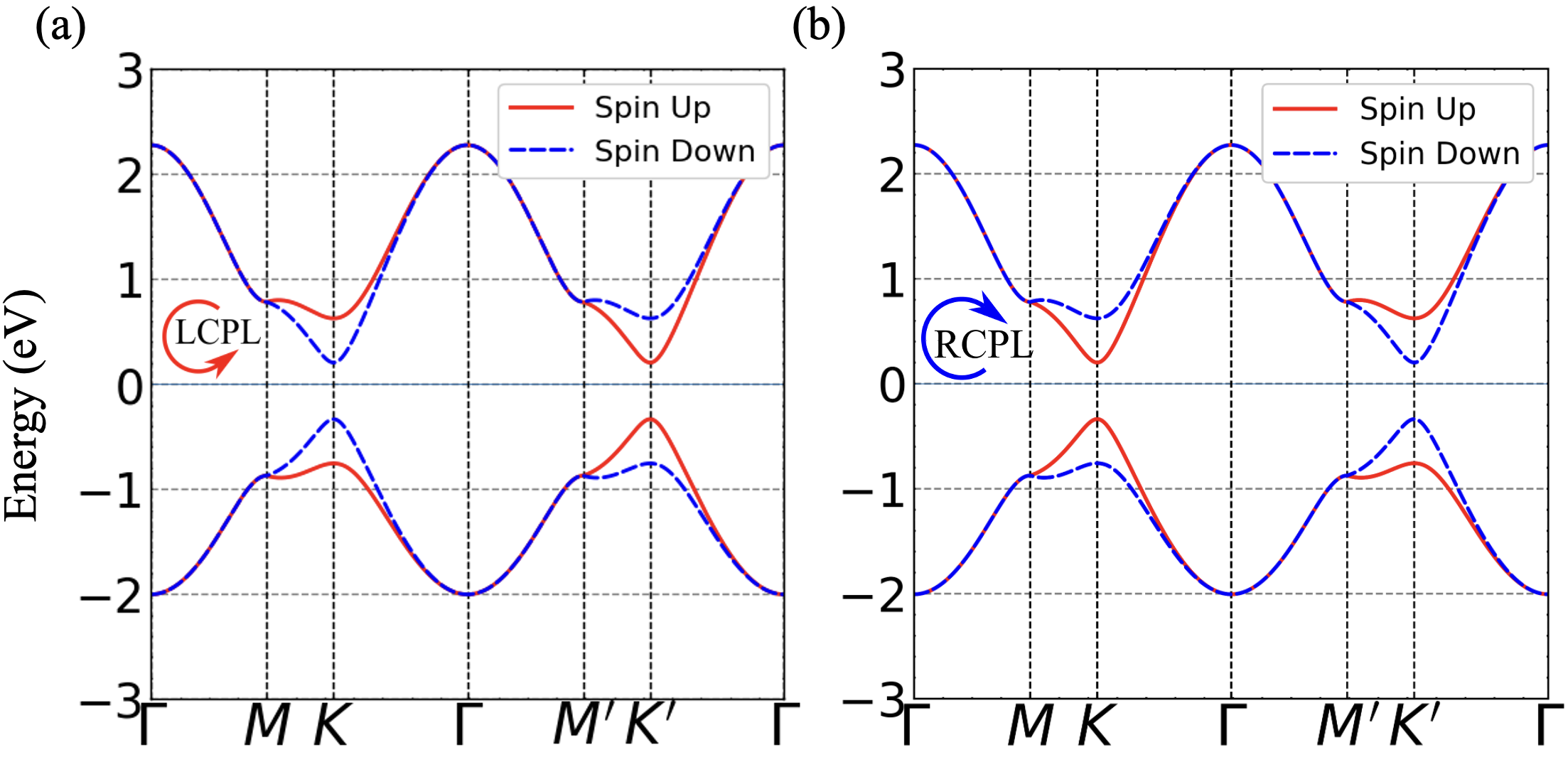}   
\caption{\label{fig:NN} The CPL irradiated band band structure for the lattice model in Eq.~(\ref{Eq12}) of the main text with additional next nearest-neighbor hopping $t_{NN}$. 
The parameters $t=1$~eV, $t_{2}=0.1$~eV, $m_1=-m_2=0.5$~eV, $\mu_{\zeta}=0$, and the light intensity of $e \mathrm{A}_0 / \hbar = 0.5~\mathring{\mathrm{A}}^{-1}$ are adopted.
}
\end{figure}

Next, we add the additional next nearest hopping term $t_{2}$ to the Hamilton as
\begin{eqnarray}\label{Eq12}
H & = & t\sum_{\left\langle i,j\right\rangle }c_{i}^{\dagger}c_{j}+ t_{2}\sum_{\left\langle\langle i,j\right\rangle\rangle}c_{i}^{\dagger}c_{j} +\sum_{i}c_{i}^{\dagger}\left[\mu_{\zeta}\sigma_{0}+\boldsymbol{m}_{\zeta}\cdot\boldsymbol{\sigma}\right]c_{i}+h.c.,
\end{eqnarray}
where $\left\langle\langle i,j\right\rangle\rangle$ is the next nearest neighbor, the Hamiltonian in momentum space can be written in the same form as 
\begin{eqnarray}\label{Eq13}
H_{\mathrm{AFM2}}\left(\boldsymbol{k}\right) & = & \sigma_0\left[\begin{array}{cc}
\Delta_{2}\left(\boldsymbol{k}\right) & \Delta\left(\boldsymbol{k}\right)\\
\Delta\left(\boldsymbol{k}\right)^{*} & \Delta_{2}\left(\boldsymbol{k}\right)
\end{array}\right]\\ \nonumber
 &  & +\left(\left|m_{\zeta}\right|\sigma_{z}+\left|\mu_{\zeta}\right|\sigma_{0}\right)\tau_{z}
\end{eqnarray}
with $ \Delta_{2}{\left(\boldsymbol{k}\right)}=t_{2}\sum_{j}e^{i\boldsymbol{k}\cdot\boldsymbol{\delta'}_{j}}$ , and $\boldsymbol{\delta'}_{j} (j=1,2,3,4,5,6)$ are the six next nearest-neighbor vectors. 
In Fig.~\ref{fig:NN}, we show that the spin-resolved band structure of left- and right-handed CPL irradiated hexangular AFM lattice with the next nearest-neighbor couplings are taken into account.
The spin-degenerate system is found to exhibit $f$-wave spin splitting under light irradiation.
In Figs.~\ref{fig:NN}(a)-(b) confirm that changing the chirality of CPL can reverse the spin polarization order.

\subsection{Odd-parity spin splitting realized in light irradiated van der Waals magnetic multilayer}

\begin{figure}[h]
\includegraphics[width=0.5\linewidth]{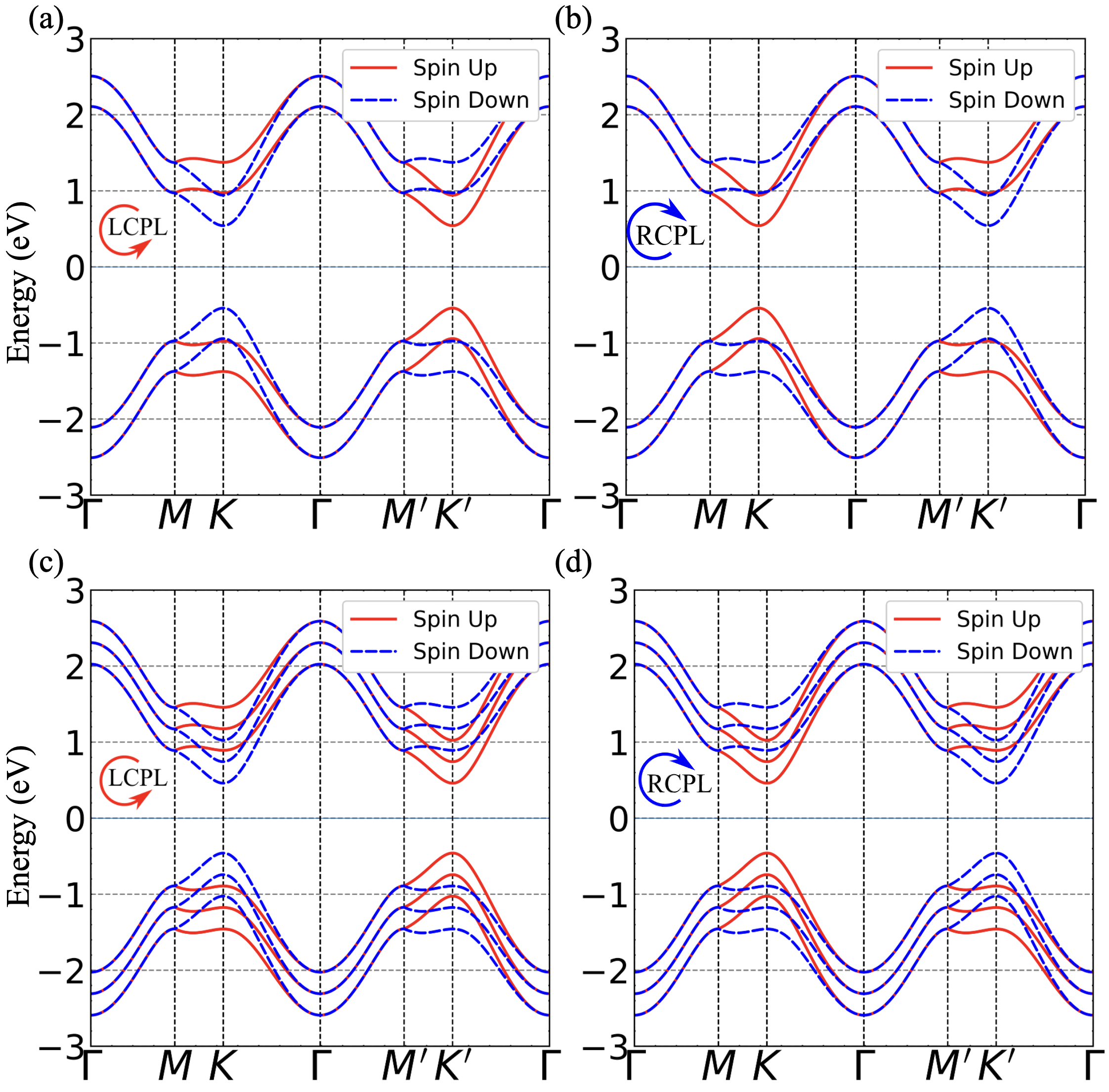}
\caption{\label{fig:ML}Odd-parity spin splitting realized in light irradiated van der Waals magnetic multilayer.
(a)-(b) The band structure of bilayer case in Eq.~(\ref{Eq14}) under right and left-handed CPL.
(c)-(d) The band structure of triplayers case in Eq.~(\ref{Eq14}) under right and left-handed CPL
the parameters $t=1$~eV, $t_{i}$=0.2~eV, $m_1=-m_2=1$~eV, $\mu_{\zeta}=0$, and the light intensity of $e \mathrm{A}_0 / \hbar = 0.5~\mathring{\mathrm{A}}^{-1}$ are adopted.
}
\end{figure}

Next, we turn to the case of AFM multilayers, by stacking the AFM model in Eq.~(\ref{Eq11}) and add inter-layer hopping $t_i$. Under the basis $\left[\varphi_{\uparrow}^{\mathrm{A1}},\varphi_{\uparrow}^{\mathrm{B1}},\cdots,\varphi_{\uparrow}^{\mathrm{A}n},\varphi_{\uparrow}^{\mathrm{B}n},\varphi_{\downarrow}^{\mathrm{A1}},\varphi_{\downarrow}^{\mathrm{B1}},\cdots,\varphi_{\downarrow}^{\mathrm{A}n},\varphi_{\downarrow}^{\mathrm{B}n}\right]$, we write the Hamiltonian in momentum space as


\begin{eqnarray}\label{Eq14}
H_{\mathrm{AFM}}^{n}\left(\boldsymbol{k}\right)=\tau_{0}I_{n}\left[\begin{array}{cc}
0 & \Delta\left(\boldsymbol{k}\right)\\
\Delta\left(\boldsymbol{k}\right)^{*} & 0
\end{array}\right]+\tau_{z}I_{n}\left(\left|m_{\zeta}\right|\sigma_{z}+\left|\mu_{\zeta}\right|\sigma_{0}\right)+t_i\tau_{0}M_{n}\sigma_{0},
\end{eqnarray}
where $I_{n}$ denotes the identity matrix of order $n$, $M_{n}$ is $n\times n$ matrix with elements read as $M_{ij}=\begin{cases}
1 & \left|i-j\right|=1\\
0 & \left|i-j\right|\neq1
\end{cases}$
, and $n$ is the number of layers. In Fig.~\ref{fig:ML}, we show that the CPL induce spin splitting in van der Waals magnetic bilayer and triple-layer systems.

\begin{figure}[h]
\includegraphics[width=0.4\linewidth]{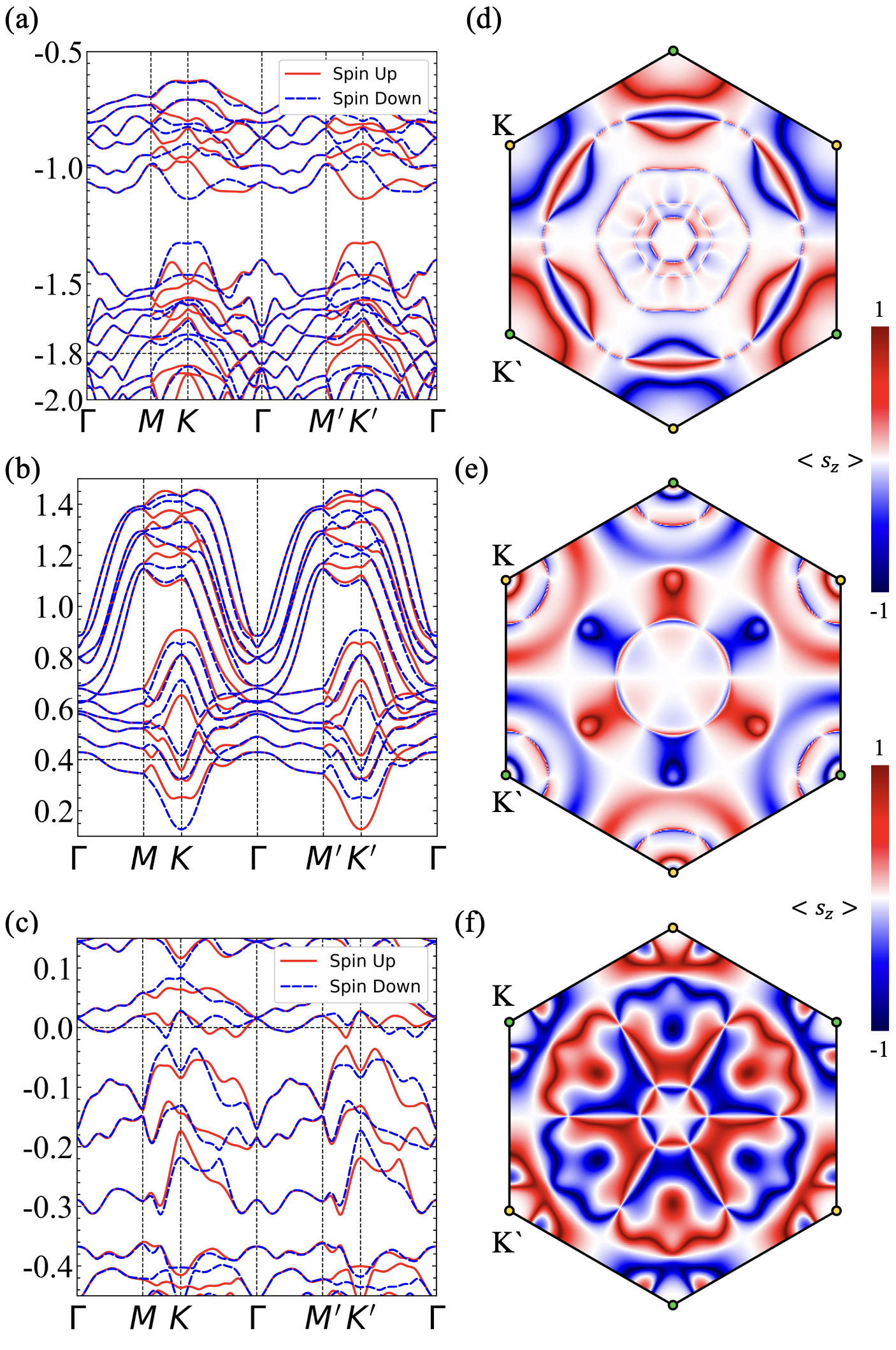}
\caption{\label{lowdft}
    low frequency CPL induced $f$-wave spin splitting in representative materials, (a)-(c) The spin-resolved band structures (curves in red solid represent the spin-up state ) along high-symmetry lines of the  AFM $\mathrm{MnPS_3}$ monolayer,  AFM $\mathrm{FeCl_2}$ bilayer, and   AFM $\mathrm{NiRuCl_6}$ bilayer under irradiation of CPL with a light intensity of $e \mathrm{A}_0 / \hbar = 0.2~\mathring{\mathrm{A}}^{-1}$ and the photon energy of this light $\hbar \omega= 0.1 $~$\text{eV}$.
    The light irradiation  preserves the spin degeneracy along the high-symmetry line $\Gamma$-M while lifting the degeneracy at  at other arbitrary $\boldsymbol{k}$-point.
    (d)-(f) The spin-resolved isoenergy surfaces with $f$-wave spin splitting of the AFM $\mathrm{MnPS_3}$ monolayer at $-1.8$~eV, the AFM $\mathrm{FeCl_2}$ bilayer at $0.4$~eV, and the AFM $\mathrm{NiRuCl_6}$ bilayer at $0.0$~eV.
}
\end{figure}

One may notes that the parameter of $\hbar\omega=10~$eV is in the high-photon energy range. 
However, we would like to  emphasize that the occurrence of the light-induced odd-parity spin splitting can be determined through symmetry analysis and is independent of the incident  photon energy. 
Therefore, when the photon energy is less than the bandwidth, light-induced odd-parity spin splitting is expected to occur.
As shown in FIG.~\ref{lowdft}, 
taking the representative material systems described by  Fig.3 of the main text as examples, under the conditions of photon energy of $\hbar\omega=0.1$~eV and light intensity of $e \mathrm{A}_0 / \hbar = 0.2~\mathring{\mathrm{A}}^{-1}$, the light irradiated antiferromagnets exhibits odd-parity spin splitting. It`s worth noting that under this low frequency range the Eq.(6) is applied with truncation at first order ($\left | \alpha-\beta \right |=0,1$).


\subsection{Model Hamiltonian of \textit{Category}-II and \textit{Category}-III }

\begin{figure}[h]
\includegraphics[width=0.4\linewidth]{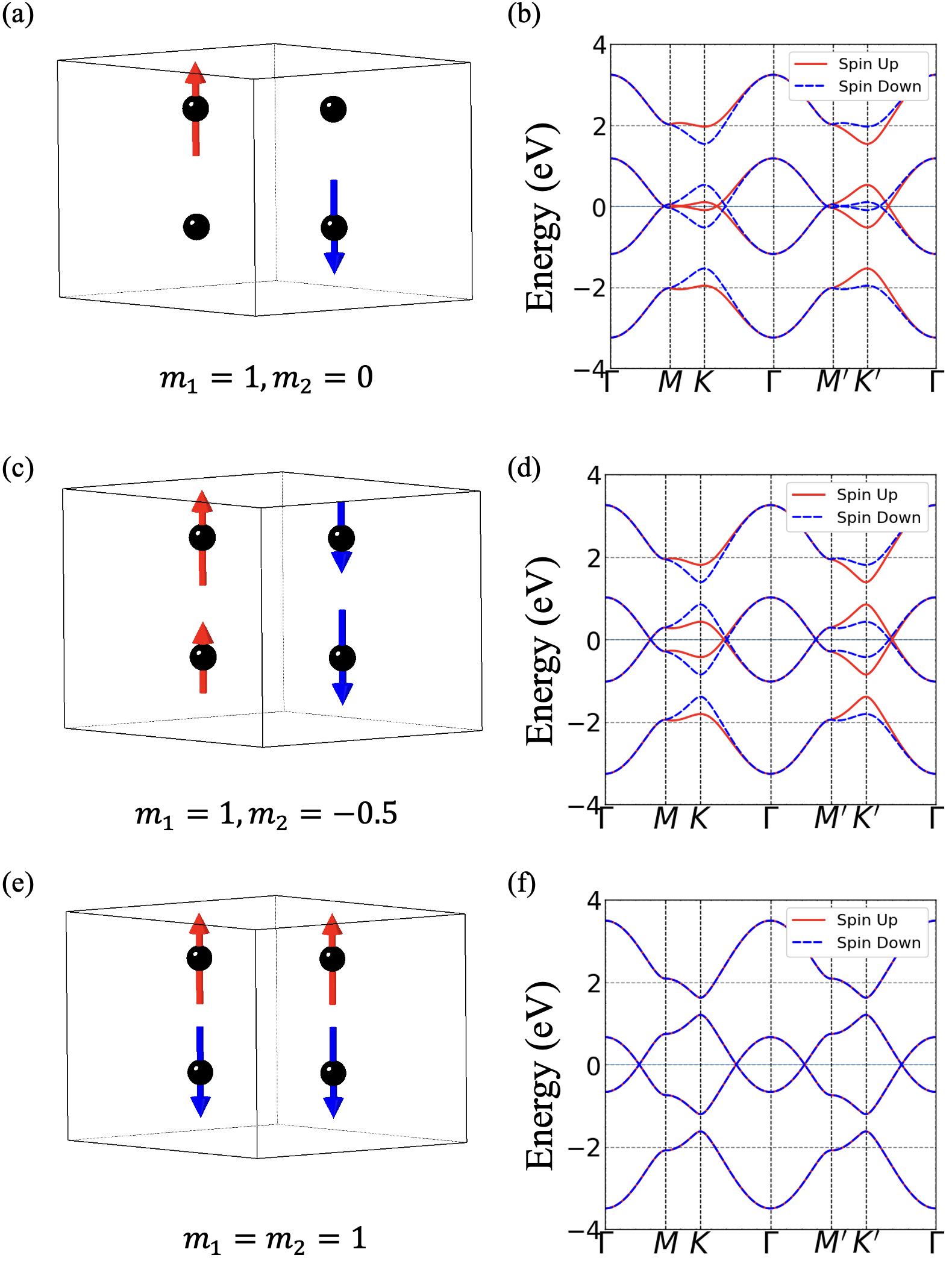}
\caption{\label{fig:23} 
The spin-splitting and degeneracy in AFM bilayers.
(a)-(b) The magnetic configuration and CPL irradiated band structure of \textit{Category}-II AFM bilayers.
(c)-(d) The magnetic configuration and CPL irradiated band structure of \textit{Category}-III AFM bilayers.
(c)-(d) The magnetic configuration and CPL irradiated band structure AFM bilayers with $\left[ \mathcal{C}_{2}\bigparallel \mathcal{M}_{z}  \right] $ symmetry.
The parameters $t=1$~eV, $t_{i}=0.1$~eV, $\mu_{\zeta}=0$, and the light intensity of $e \mathrm{A}_0 / \hbar = 0.5~\mathring{\mathrm{A}}^{-1}$ are adopted.
}
\end{figure}

Next we turn to the AFM bilayers composed of magnatic monolayers which is referred as \textit{Category}-II and \textit{Category}-III in the main text. Under the basis $\left[\varphi_{\uparrow}^{\mathrm{A1}},\varphi_{\uparrow}^{\mathrm{B1}},\varphi_{\uparrow}^{\mathrm{A}2},\varphi_{\uparrow}^{\mathrm{B}2},\varphi_{\downarrow}^{\mathrm{A1}},\varphi_{\downarrow}^{\mathrm{B1}},\varphi_{\downarrow}^{\mathrm{A}2},\varphi_{\downarrow}^{\mathrm{B}2}\right]$,
we write the Hamiltonian in momentum space as
\begin{eqnarray}\label{Eq5}
H_{\mathrm{AFM}}^{n}\left(\boldsymbol{k}\right)=\tau_{0}I_{2}\left[\begin{array}{cc}
0 & \Delta\left(\boldsymbol{k}\right)\\
\Delta\left(\boldsymbol{k}\right)^{*} & 0
\end{array}\right]+\tau_{z}\left(iM'\left|m_{\zeta}\right|\sigma_{z}+I_2\left|\mu_{\zeta}\right|\sigma_{0}\right)+t_i\tau_{0}M_{2}\sigma_{0},
\end{eqnarray}
where $M'$ is the second Pauli matrix indicates the interlayer AFM order. 
When $m_{1}\ne0$ $\&$ $m_{2}=0$ ($m_{1}\ne m_{2}\ne0$), the van der Waals magnetic bilayer system is classified as \textit{Category}-II (\textit{Category}-III) in the main text. The CPL induced $f$-wave spin splitting in these two kinds of lattice model are shown in  Fig.~\ref{fig:23}. Notably, when $m_{1}=m_{2}=0$ the two sublattice can be connected by $\left[ \mathcal{C}_{2}\bigparallel \mathcal{M}_{z}  \right] $, indicating the absence of spin splitting as shown in Figs.~\ref{fig:23} (e)-(f).


\section{The spin degeneracy in intrinsic AFM $\mathrm{MnPS_3}$, $\mathrm{FeCl_2}$, and $\mathrm{NiRuCl_6}$ }

\begin{figure}
\includegraphics[width=0.7\linewidth]{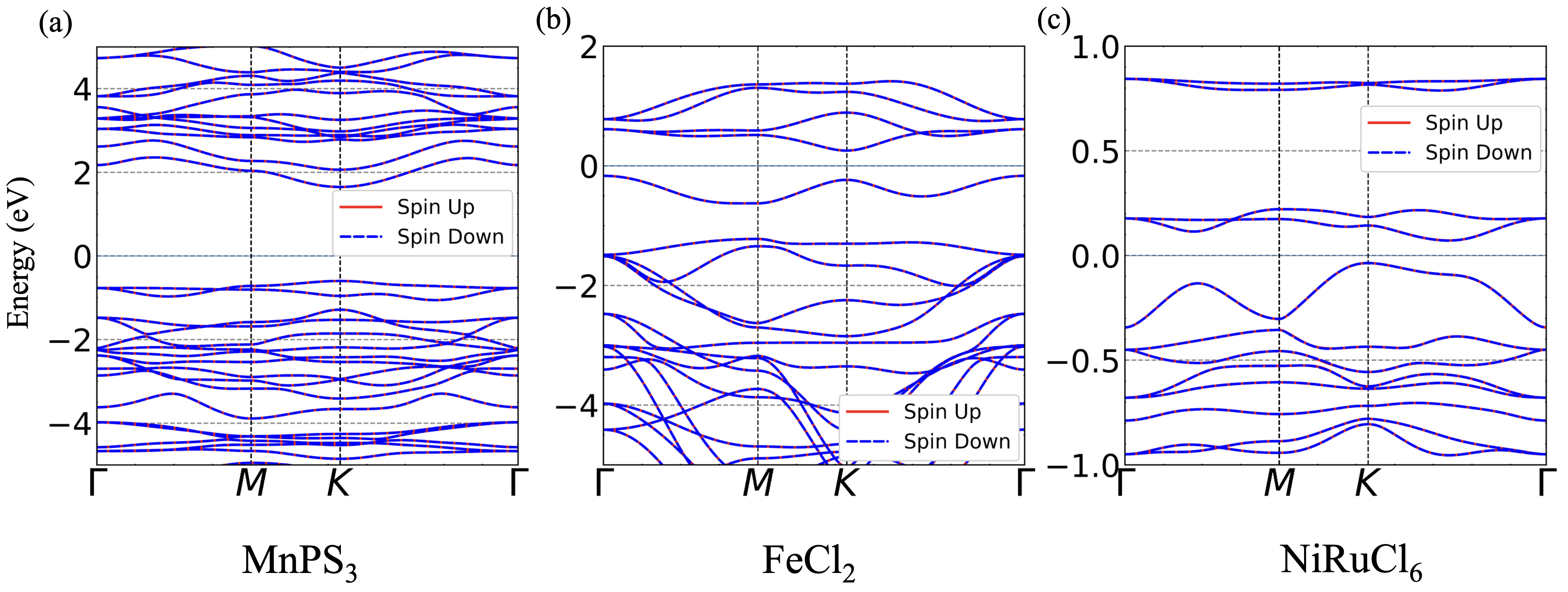}
\caption{\label{fig:AFM} The intrinsic spin-degenerate band structure of (a) $\mathrm{MnPS_3}$ monolayer, (b) AFM $\mathrm{FeCl_2}$ bilayer, and (c) AFM $\mathrm{NiRuCl_6}$ bilayer without spin-orbital coupling.
}
\end{figure}

Fig.~\ref{fig:AFM} demonstrates that the energy bands of these three materials are intrinsically spin-degenerate in the absence of a light field and spin-orbital coupling.

\section{The spin splitting in BCL irradiated conventional  antiferromagnetism}
\begin{figure}
\includegraphics[width=0.56\linewidth]{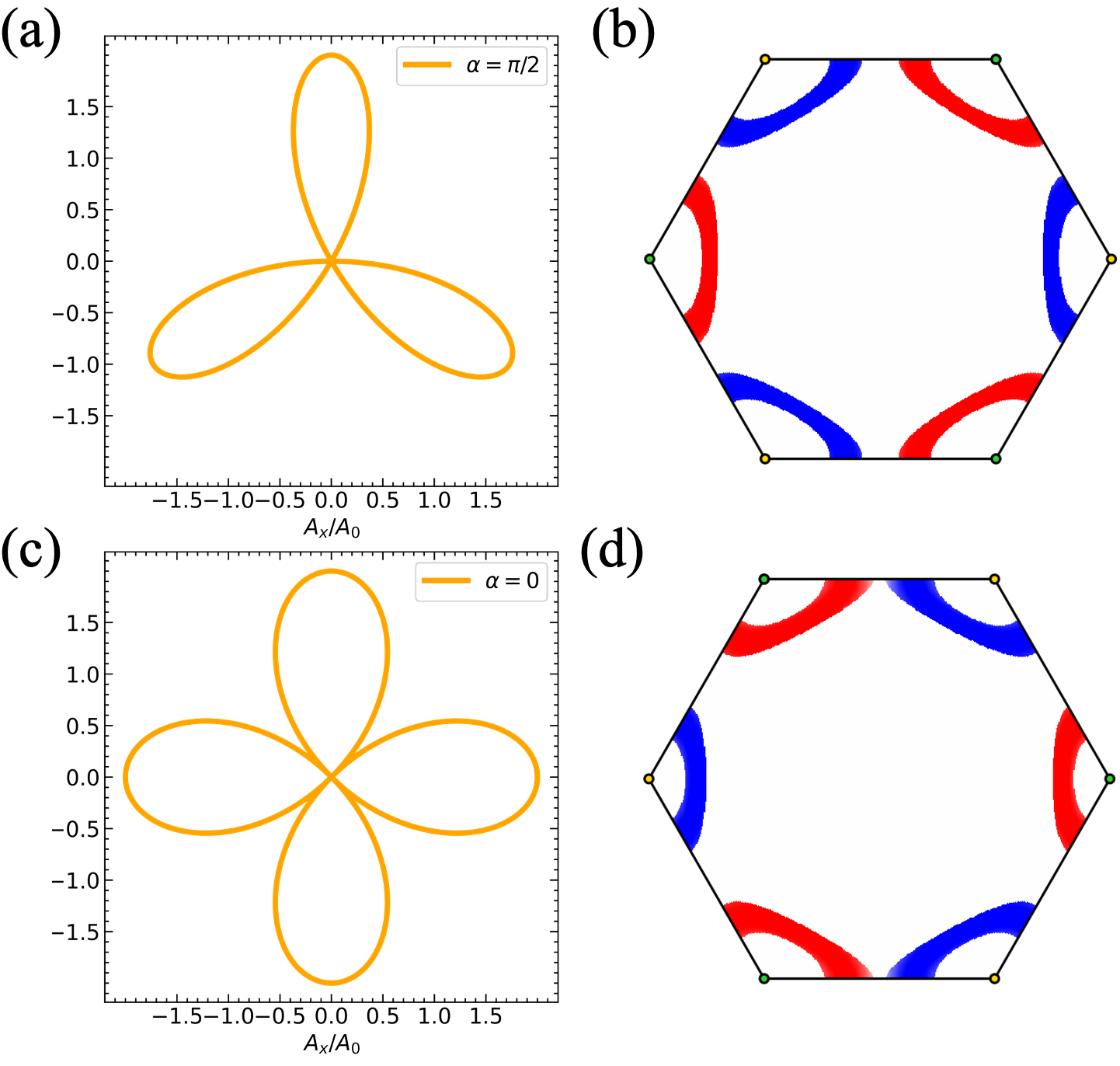}
\caption{\label{fig:add1}
The minimum hexagonal lattice model with conventional AFM order (Fig.2 in the main text) under BCL with clover pattern and 4-leaf clover pattern.
(a) The light polarization pattern for the BCL of $\kappa=2$. 
(b) The spin-resolved isoenergy surface at -1 eV.
(c) The light polarization pattern for the BCL of $\kappa=3$. 
(d) The spin-resolved isoenergy surfaces at -1 eV. 
In panels (b) and (d), the parameters $t=t_{2,3}=m_1=-m_2=1$~eV and the light intensity of $e \mathrm{A}_0 / \hbar = 0.5~\mathring{\mathrm{A}}^{-1}$ are adopted.
}
\end{figure}

In this section, we discuss the spin splitting induced by  BCL with the minimum lattice model in the main text. 
The vector potential of BCL is  expressed as
\begin{equation}\label{Eq.add1}
\mathbf{A}(t)={A}_0\sqrt{2}\mathrm{Re}[e^{-i(\kappa\omega t-\alpha)}\boldsymbol{\varepsilon}_R+e^{-i\omega t}\boldsymbol{\varepsilon}_L],
\end{equation}
where $A_0$ is amplitude for both right-handed circularly polarized (RCP) and right-handed circularly polarized (LCP) light, $\varepsilon_{R(L)}$ is RCP (LCP) light polarization basis vectors, $\kappa$ and $\alpha$ are frequency ratio and phase difference between RCP and LCP light respectively. 
For collinear spin configurations, the spin-only group is given by $\mathbf{r}_{s}=\mathbf{C}_{\infty}+\bar{C}_{2}\mathbf{C}_{\infty}$, where $\mathbf{C}_{\infty}$ denotes arbitrary spin rotations around the common spin axis~\cite{rn1l-d6cq}. 
While circularly polarized light breaks $\bar{C}_{2}$ and preserves $\mathbf{C}_{\infty}$, bicircularly polarized light breaks  $\bar{C}_{2}$ and reduces  $\mathbf{C}_{\infty}$ to a specific rotational symmetry determined by the ratio of the light frequencies.
As show in Fig.~\ref{fig:add1} (a) and (c), when $\kappa=2$ and $\kappa=3$,  the symmetry of the light polarization pattern is reduced to $C_3$ and $C_4$, respectively. 
Fig.~\ref{fig:add1} (b) and (d) indicate that when the $C_3$ ($C_4$) symmetry is preserved in BCL, one can reach $f$-wave ($p$-wave) spin-splitting.

\section{The $p$-wave spin splitting in $\mathrm{FeCl_2}$ induced by BCL, EPL, and strain}

\begin{figure}[h]
\includegraphics[width=0.55\linewidth]{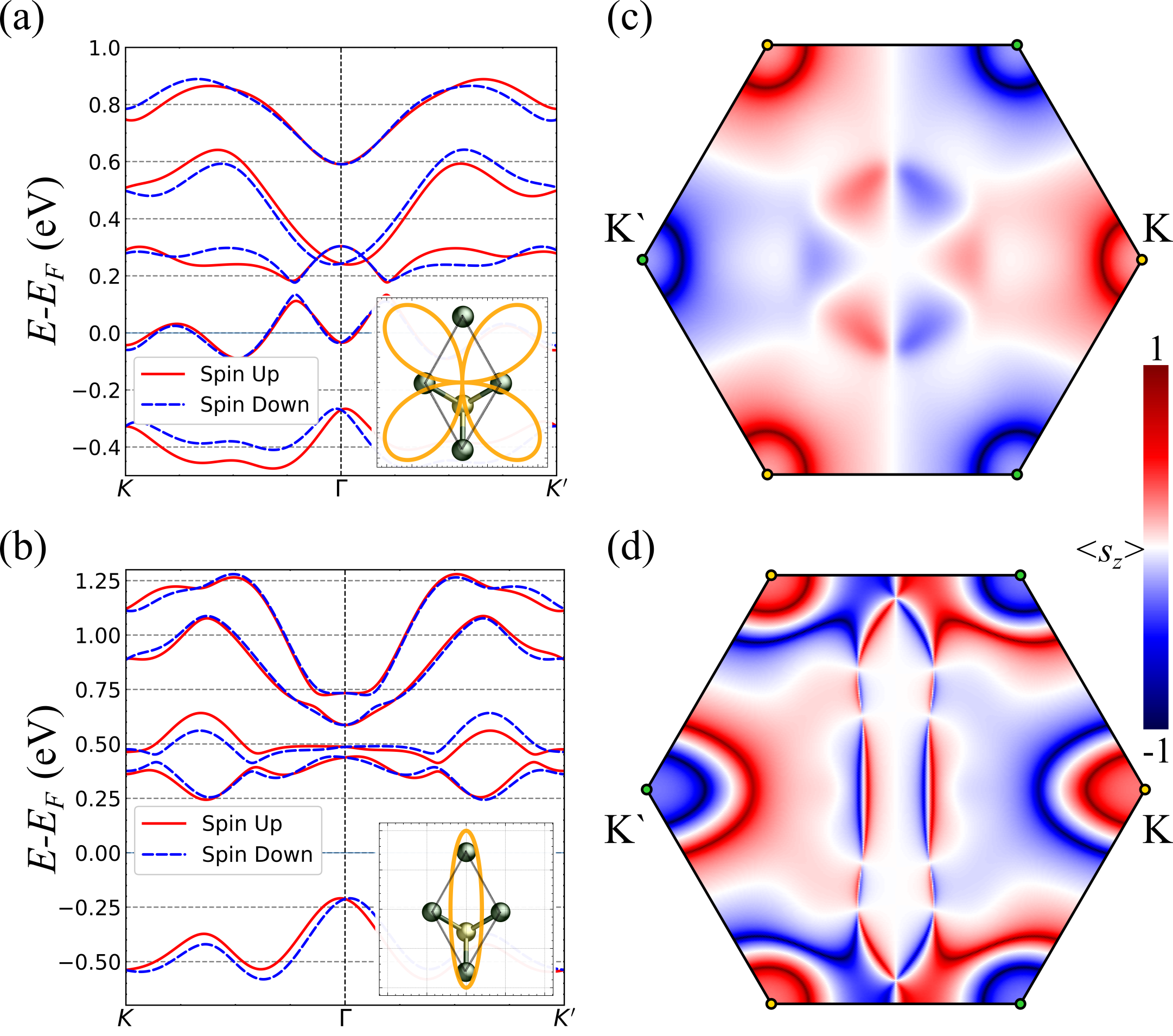}
\caption{\label{fig:LPLM} BCL and EPL induced $p$-wave spin-splitting for $\mathrm{FeCl_2}$ bilayer .
(a)-(b) The band along high-symmetry lines of the $\mathrm{FeCl_2}$ bilayer under BCL and EPL with a light intensity of $e \mathrm{A}_0 / \hbar = 0.3$~\AA$^{-1}$. 
The BCL is as descried in \ref{Eq.add1}, with $\kappa=3$ and $\phi=0$
The insert in each figure shown the relative orientation of light polarization pattern and $\mathrm{FeCl_2}$ lattice.
(c)-(d) The spin-resolved p-wave isoenergy surface at 0.5 eV of $\mathrm{FeCl_2}$ bilayer corresponding to (a) and (b)
}
\end{figure}

\begin{figure}[h]
\includegraphics[width=0.55\linewidth]{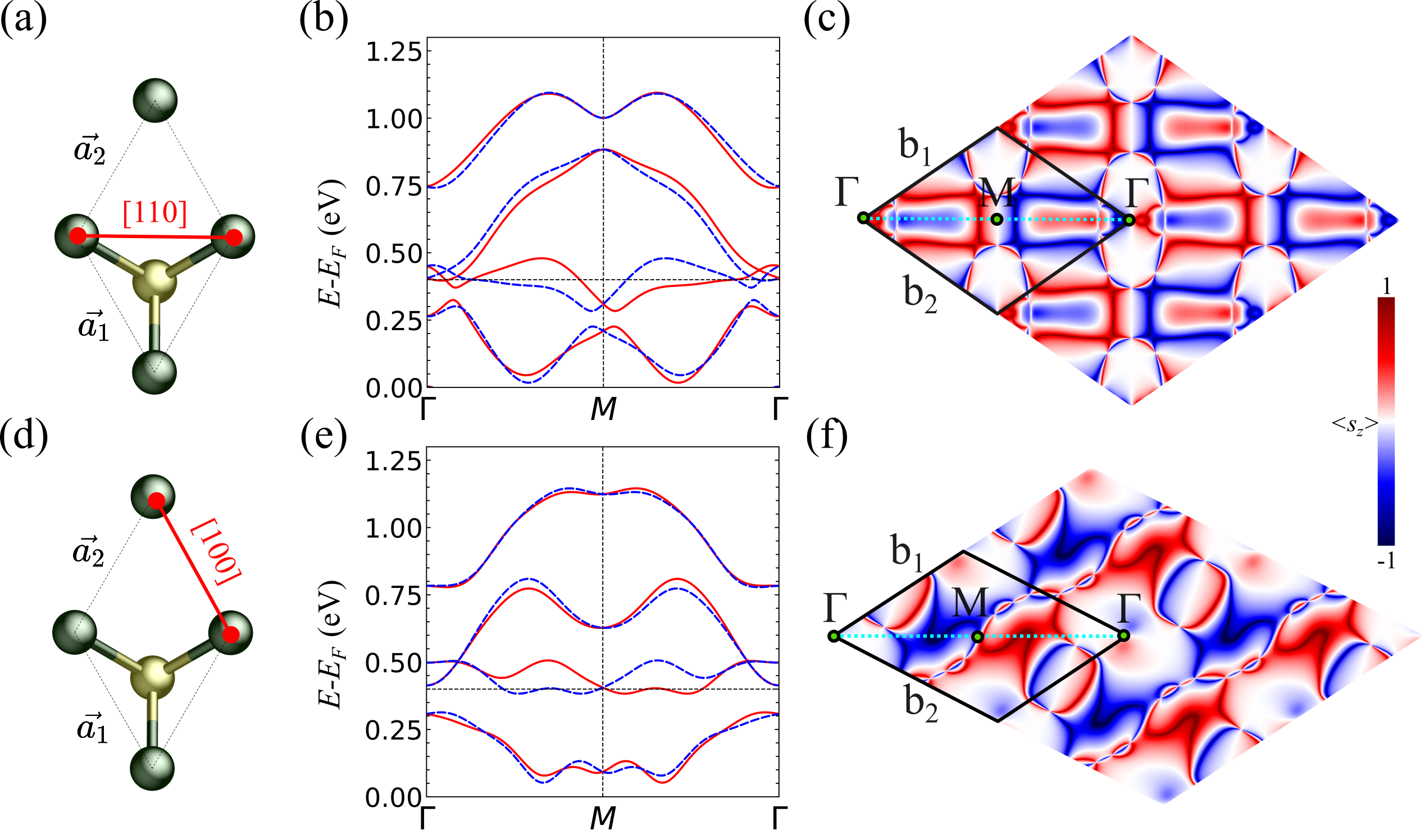}
\caption{\label{fig:LPLM1}CPL induced $p$-wave spin-splitting for strain applied $\mathrm{FeCl_2}$ bilayer .
(a)-(c) CPL irradiated $\mathrm{FeCl_2}$ bilayer with 20\% tensile strain along [110] direction.
(d)-(f) CPL irradiated $\mathrm{FeCl_2}$ bilayer with 20\% tensile strain along [100] direction.
(a) and (d) are the illustration figure of corresponding strain. 
In (b) and (a) the spin-resolved band structures are shown for both cases.
(c) and (f) are The spin-resolved $p$-wave isoenergy surface at 0.4 eV for both cases. A CPL with the light intensity of $e \mathrm{A}_0 / \hbar = 0.3$~\AA$^{-1}$ are adopted. 
}
\end{figure}

In Figs.~\ref{fig:LPLM}-\ref{fig:LPLM1}, we take the AFM $\mathrm{FeCl_2}$ bilayer as an example to show that the BCL (which is as descried in Eq.\ref{Eq.add1}, with frequency ratio $\kappa=3$ and $\alpha=0$), EPL, and strain combined with CPL can induce $p$-wave spin splitting.

\newpage
\newpage
\section{material candidates}
In this section, we list some material candidates for light-induced $f$-wave spin-splitting. All of the materials listed in Table~\ref{tab:table1} have been experimentally or computationally reported to have magnetic order that required to realize odd-parity spin splitting.
\begin{table*}[h]
\caption{\label{tab:table1}%
The 2D material candidates for odd-parity magnetism and Floquet induced odd-parity magnetism. The space group and Transition temperatures of these material candidates are tabulated.}
\begin{ruledtabular}
\begin{tabular}{cccc}
Type&Material&Space group &Transition temperature\\
\hline
Noncollinear &$\mathrm{NiI_2}$\cite{oddNatruesong2025electrical} & $R3\bar{m}$ & 59.5K\\
\hline
\textit{Category}-\MakeUppercase{\romannumeral 1} &$\mathrm{MnPS_3}$\cite{MnPS,MnPS2} & $P\text{-}31m$ & 78K\\
\textit{Category}-\MakeUppercase{\romannumeral 1} &$\mathrm{MnPSe_3}$\cite{MnPS2,addtc1} & $P\text{-}31m$& 74K\\
\textit{Category}-\MakeUppercase{\romannumeral 1} &$\mathrm{Fe_2O_3}$\cite{FeO} & $P\text{-}31m$ & 270K (Calculation)\\
\textit{Category}-\MakeUppercase{\romannumeral 1} &$\mathrm{MnX (X=S,Se,Te)}$\cite{MnX} & $P\text{-}3m1$& - \\
\textit{Category}-\MakeUppercase{\romannumeral 1} &$\mathrm{Fe_2C}$\cite{FeC} & $P\text{-}31m$& 390K (Calculation)\\
\textit{Category}-\MakeUppercase{\romannumeral 1} &$\mathrm{Cr_2CCl_2}$\cite{CrCCl} & $P\text{-}3m1$&-\\
\textit{Category}-\MakeUppercase{\romannumeral 1} &$\mathrm{MnPTe_3}$\cite{MnPTe} & $P\text{-}31m$&-\\
\textit{Category}-\MakeUppercase{\romannumeral 1} & BL-$\mathrm{MnBi_2Te_4}$\cite{MnBiTe,MnBiTe2} & $P\text{-}3m1$&25K\\
\textit{Category}-\MakeUppercase{\romannumeral 2} &1H-$\mathrm{RuO_2}$\cite{RuO} &  $P\bar{6}m2$&-\\
\textit{Category}-\MakeUppercase{\romannumeral 2} &1H-$\mathrm{MX_2(M=V,Nb,Ta;X=S,Se,Te)}$\cite{VTe} &  $P\bar{6}m2$&-\\
\textit{Category}-\MakeUppercase{\romannumeral 2} &1H-$\mathrm{FeX_2(X=O,S,Se,Te)}$\cite{FeX} &  $P\bar{6}m2$&-\\
\textit{Category}-\MakeUppercase{\romannumeral 2} &1H-$\mathrm{MnTe_2}$\cite{MnXX,MnXX2} &  $P\bar{6}m2$&-\\
\textit{Category}-\MakeUppercase{\romannumeral 2}&1H-$\mathrm{MnO_2}$\cite{MnXX,MnXX2} &  $P\bar{6}m2$&140K (Calculation)\\
\textit{Category}-\MakeUppercase{\romannumeral 3} & $\mathrm{NiRuCl_6}$\cite{NiRuCl,NiRuCl2} & $P321$&-\\
\end{tabular}
\end{ruledtabular}
\end{table*}

\section{A brief discussion on experimental realization of light-induced odd parity spin splitting}

In this section, we would like to give a brief discussion on experimental realization of light-induced odd parity spin splitting.
It is well know that to detect the light driving states experimentally is undoubtedly a challenging endeavor. For example, under intense light irradiation, the system may experience heating, which could potentially suppress the magnetic order if the effective temperature exceeds the Néel temperature. However, we also believe that it is most encouraging to see experimental attempts being made on this issue. We would like to addressit as follows: I. \textit{The transition temperature can be relatively high}:
As tabulated in TABLE~\ref{tab:table1}, The transition temperatures of NiI$_2$ is 59.5K. However, the transition temperatures of the collinear candidates range from 74K to 390K (except for MnBi$_2$Te$_4$). The higher transition temperatures make collinear candidates better for real applications. 
II. \textit{Heating suppression in the high-frequency regime:}
Our work can be applied to the high-frequency, off-resonant regime,
where recent theoretical studies have shown that heating in periodically driven systems can be exponentially suppressed \cite{RevModPhys.93.041002,PhysRevLett.116.120401}.
This suggests that the predicted light-induced spin-splitting states here may have sufficiently long lifetimes to be observable.
III. \textit{ Floquet states and spin splitting are experimentally detectable}: The Floquet states can be detected by advanced time-resolved techniques, such as time-resolved and angle-resolved photoemission spectroscopy (TrARPES) and time-resolved transport measurements \cite{zhou2023pseudospin,ito2023build}. 
The spin-resolved TrARPES is also thriving on experimental side. As the section VII of the Review of Morden Physics paper \cite{RevModPhys.96.015003} described, the spin-resolved TrARPES has been used in materials like ferromagnets, topology insulators, topology semimetals, and getting the spin-resolved ARPES in ultrafast timescale.
IV. \textit{ Unique transport properties}:  The odd-parity spin-splitting can contribute distinctive transport properties such as anisotropic bulk spin conductivity\cite{zeng2025electronic}, large tunneling magnetoresistance\cite{PhysRevLett.133.236703}, nonrelativistic Edelstein response \cite{yu2025oddparitymagnetismdrivenantiferromagnetic}
as proposed. Thus, the light-induced odd parity spin splitting  can be detected by ultrafast time-resolved transport measurements.
V. \textit{ Robustness}: Moreover the symmetry manipulation mechanism we proposed can be applied in many materials under various light parameters, which offers a broader possibility for experimental realization.
VI. \textit{Larger range of material candidates}: As tabulated in TABLE S1 in the SM, there are many 2D collinear AFMs that are material candidates for light-induced $f$-wave spin splitting. Nevertheless, to best of our knowledge,  spin splitting in 2D non-collinear AFMs has only been reported in NiI$_2$. Thus, the range of available collinear candidate materials is substantially larger than that of non-collinear candidates. In particular,  the collinear candidates tabulated in TABLE S1 are well studied. We believe the collinear AFMs provide us an ideal platform to research odd-parity magnetism.
VII. \textit{The collinear candidates are much more accessible both theoretically and experimentally.} 
Theoretically, collinear AFMs often preserve higher symmetry and fit neatly into standard magnetic space groups (Shubnikov groups), while the magnetic unit cell of non-collinear AFMs become very large or incommensurate, such as in helical or cycloidal structures, making theoretical modeling of the supercell extremely difficult. Experimentally, collinear magnets produce straightforward magnetic Bragg peaks in neutron diffraction that are easier to index and refine. In contrast, non-collinear structures often generate weak satellite peaks or ambiguous interference patterns that are notoriously difficult to distinguish from multi-domain averaging effects.



\end{document}